\documentclass{article}       

 \usepackage{graphicx}
 \usepackage{mathptmx}      
 \usepackage{amsmath,comment,fancyhdr,color,graphics}
 \usepackage{amsfonts}
 \usepackage{natbib}
 \usepackage{xcolor}

\newcommand{\eps}{\epsilon}
\newcommand{\beq}{\begin{equation}}
\newcommand{\enq}{\end{equation}}
\newcommand{\beqa}{\begin{eqnarray}}
\newcommand{\enqa}{\end{eqnarray}}
\newcommand{\neno}{\newline\noindent}
\newcommand{\tg}{\tilde{g}}
\newcommand{\esp}{\mathbb{E}}
\renewcommand{\H}{{\cal H}}
\newcommand{\F}{{\cal F}}
\newcommand{\G}{{\cal G}}
\newcommand{\lx}{{\cal L}(X)}
\newcommand{\ly}{{\cal L}(Y)}
\newcommand{\vg}{\mathbf{g}}
\newcommand{\tvg}{\tilde{\mathbf{g}}}
\newcommand{\cls}{{\cal L^*}}

\newcommand{\X}{{\cal X}}
\newcommand{\ip}{\mbox{\small(p)}}
\newcommand{\cip}{\stackrel{\ip}{\to}}
\newcommand{\dife}{\mbox{d}\,}
\newcommand{\dd}{\mbox{d}}

\newtheorem{prop}{Proposition}

\begin{document}

\title{Quadratic forms of the empirical processes for the two sample
problem for functional data}

\author{R. Barcenas$^a$ \hspace{1cm}
        J. Ortega$^a$   \hspace{1cm}
        A. J. Quiroz$^b$ \\  \\
        $^a$ {\small \textit{Dpto. de Probabilidad y Estad\'{\i}stica.
        CIMAT, A.C.}}\\{\small \textit{ Jalisco, s/n, Mineral de Valenciana.
        Guanajuato 36240, Mexico.}}\\
        $^b$ {\small \textit{Dpto. de
    Matem\'aticas, Universidad de Los Andes.}}\\ {\small \textit{Carrera 1, Nro. 18A-10,
    edificio H, Bogot\'a, Colombia.}}\\ {\small \textit{Phone: (571)3394949, ext. 2710. Fax:
    (571)3324427.}}
        }
\date{}

%
%

\maketitle

\begin{abstract}
The use of quadratic forms of the empirical process for the
two-sample problem in the context of functional data is considered.
The convergence of the family of statistics proposed to a Gaussian
limit is established under metric entropy conditions for smooth
functional data. The applicability of the proposed methodology is
evaluated in examples.

\end{abstract}


\section{Introduction}
\label{intro}

Functional data analysis has had a very important growth in the last
20 years, and has found applications in many different areas,
especially since the first edition of the book by Ramsay and
Silverman in 1997. More recent contributions to the field can be
found in \citet{bosq}, \citet{rs1, rs2}, \citet{fv}, \citet{ferraty}
 and \citet{hk}, where examples of diverse applications can also be
found.

In the analysis of functional data a frequent problem is that of
deciding if two samples of functions come from the same population.
Let $X_1(t), \cdots, X_m(t)$ be an i.i.d. sample of real valued curves defined
on some interval $J$. Denote by
$\mathcal L(X)$ the probability law producing these curves. Likewise, let
$Y_1(t), \cdots , Y_n(t)$, be another i.i.d. sample of curves, independent
of the $X$ sample and also defined on $J$, with
probability law $\mathcal L(Y)$. In the two-sample problem, we wish to test
the null hypothesis, $H_0$: $\mathcal L(X) = \mathcal L(Y)$ against the general
alternative $\mathcal L(X) \neq \mathcal L(Y)$.

This problem has been considered from several viewpoints.
\citet{munoz} define a similarity index for curves based on the
sample correlation coefficient of vectors obtained from evaluating
the registered curves on a common grid and use permutation tests.
\citet{hvk}  study the effect of smoothing the functional data on
the power of tests for the two-sample problem and propose bootstrap
statistics that generalize the two-sample Cram\'er-von Mises
methodology to the functional data setting. \citet{benko} consider
the problem from the point of view of functional principal
components. To test the differences between two samples of functions
their respective Karhunen-Lo\`eve expansions are considered. In
particular, they develop a bootstrap test for testing common
principal components. \citet{hk2} consider the two sample problem
for regressions of the form $Y^j = \psi^j X^j+\varepsilon^j, j=1,2$,
where the $X^j$ are function over a compact subset of a Euclidean
space, the responses $Y^j$ can either be functions or scalars and
the $\psi^j$ are linear operators over a function space which take
either values in the same function space or scalar values. Using
expansions with respect to the functional principal components, they
develop a test for the equality of the operators $\psi^j$.
\citet{pena} presents several proposals for the functional
two-sample problem, based mainly on permutation tests.
 \citet{ps} develop a general
testing methodology for functional data based on bootstrap
techniques, which is applicable to different testing problems and
test statistics, including the comparison of mean or covariance
functions.

In their book, \citet[Ch. 5]{hk} consider samples $X_i^j, i=1,2,
\dots, n_j,  j=1,2$. Under the assumption that they satisfy the
models
\begin{equation}\label{ec1a}
X_i^j(t) = \mu^j(t) + \varepsilon_i^j(t),\quad 1\leq i \leq n_j,\
j=1,2
\end{equation}
they propose two tests for the hypothesis $H_0: \mu^1 = \mu^2$ in
$L^2$ against the alternative that $H_0$ is false. The first method
is based on the sample estimators for the mean functions, while the
second is based on the functional principal component expansions. We
describe the latter in detail, since it is related to the method
proposed in the present work.

Assume that the two samples are independent, the noises are
centered,  $\varepsilon_i^j, i=1,\dots, n_j$ are i.i.d. for fixed
$j$ and are independent for different $j$, although they are not
assumed to have the same distribution in this case. Also
$E||\varepsilon_1^j||^4 < \infty$ for $j=1,2$. Consider the operator
$Z=(1-\theta)C^1 + \theta C^2, 0\leq \theta, \leq 1$, where $C^j$ is
the covariance operator corresponding to $X^j, j=1, 2$. Assume the
eigenvalues of $Z$ satisfy
\begin{equation}\label{ec2a}
\tau_1 > \tau_2 > \cdots > \tau_d > \tau_{d+1}
\end{equation}
for some large $d$ and let $\phi_1,\dots, \phi_d$ be the
corresponding eigenfunctions. Let $\hat Z_{n_1,n_2}$ be the sample
version of $Z$ and let $\hat \phi_1,\dots, \hat\phi_d$ be the
eigenfunctions for this operator. Let $\bar X^j = (1/n_j)
\sum_{i=1}^{n_j} X_i^j(t)$, $\hat a_i = \langle \bar X^1-\bar
X^2,\hat \phi_i\rangle, 1\leq i \leq d$, and $\hat{\mathbf{a}} =
(\hat a_1,\dots , \hat a_d)^T$. Assume that
\begin{equation}\label{ec3a}
\frac{n_1}{n_1+n_2} \to \theta,\quad \text{for some } 0\leq
\theta\leq 1.
\end{equation}
Then, under all these conditions Horv\'ath and Kokoszka prove that
\begin{equation}\label{ec4a}
\Big( \frac{n_1n_2}{n_1+n_2} \Big)^{1/2} \hat{\mathbf{a}}\quad
\stackrel d \longrightarrow \quad N_d(\mathbf{0}, Q)
\end{equation}
where the limit is a $d$-dimensional centered normal distribution
with diagonal covariance  matrix satisfying $Q(i,i)=\tau_i$. In
consequence they propose the statistics
\begin{equation}\label{ec5a}
T^1_{n_1, n_2} = \frac{n_1 n_2}{n_1 + n_2} \sum_{k=1}^d \hat a_k^2 /
\tau_k \quad\stackrel d \longrightarrow \quad \chi^2_d
\end{equation}
and
\begin{equation}\label{ec6a}
T^2_{n_1, n_2} = \frac{n_1 n_2}{n_1 + n_2} \sum_{k=1}^d \hat a_k^2
\quad \stackrel d \longrightarrow \quad \sum_{k=1}^d \tau_k N_k^2,
\end{equation}
where $N_1,\dots, N_d$ are independent Gaussian standard random
variables.

In this work a family of statistics for the two-sample problem on
functional data is studied. It is a family of quadratic forms
associated to dot products of functions of the samples with a finite
number of adequately chosen functions. Details will be given in the
next section. This family includes $T^1$ as a special case.

As examples of applications of the family of statistics proposed here to real
data, some problems in Oceanography are considered. The stochastic
approach to the analysis of ocean waves originated in the 1950's
with the work of \citet{pi} and \citet{lh1, lh2}. This approach
considers ocean waves as a realization of a random process,
frequently a centered stationary Gaussian processes, and this point
of view has permitted the analysis of many important features of
waves. An account of this theory can be found in \citet{ochi}.

The assumption of stationarity permits the use of spectral analysis
techniques to study the wave energy distribution in the frequency
domain. This analysis is related to several important
characteristics in Oceanography, such as the significant wave height
$H_s$, (see section \ref{sec4} for a definition), a standard measure
of sea severity which can be obtained from the spectral distribution
of the process. On the other hand, Gaussian processes provide
tractable models for which it is possible to obtain explicit
distributions of many parameters of interest, and are suitable
models in many circumstances. They also provide a good first order
approximation when nonlinearities are present.

However, both hypotheses have limitations. Stationarity is only a
valid hypothesis for short periods of time, while normality fails
for shallow water waves or when nonlinearities are present. As a
first application of the class of statistics proposed, we consider
the problem of testing whether two samples of estimated spectral
densities coming from (simulated) random wave processes have the
same distribution. This is related to the problem of determining
stationary intervals in the sea surface behavior.

As regards the assumption of normality,  the Gaussian model cannot
account for observed asymmetries in real waves, a fact that has been
known for a long time. According to \citet{borgman}, \lq Gaussian
models involving superposition of linear waves predict all the
probability properties of the sea surface. Yet the commonly observed
property that wave crests reach higher above mean water level than
the troughs fall below cannot be encompassed within the model.'

\begin{figure}[t]
\centering
  \includegraphics*[bb= 5cm 20cm 18cm 26cm ,clip, width=10cm]{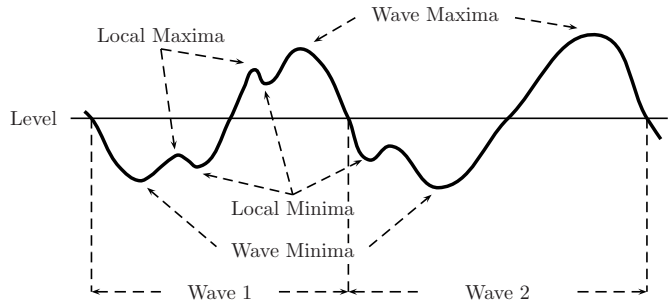}\\
  \caption{Wave characteristics}\label{fig02}
\end{figure}

In \citet{goros}, one-dimensional random waves from a North-Sea
storm were considered from a functional point of view. A wave is
defined as the trajectory of the sea-surface elevation between two
consecutive downcrossings of the mean sea level (see Figure
\ref{fig02}). The mean waves obtained after registration for a
series of 20-minute intervals were considered and several features
such as first and second derivatives, phase diagrams and their
relation with the significant wave height of the corresponding
20-minute period were analyzed. Also, a comparison between real and
simulated Gaussian waves was made. For this comparison, the spectral
density for each 20-minute period was estimated, and a Gaussian
process with the same sampling frequency as the original data was
simulated from the spectral density. Mean waves for both cases  were
compared using a randomization conditional test and the results gave
strong evidence that real and simulated waves follow different
distributions. That study also gave evidence of the asymmetry of
real waves, compared to simulated Gaussian waves.

As further applications of the family of statistics developed in
this work, we consider two problems associated to the analysis of
random waves as functional data. The first concerns the effect that
the amount of energy present in the sea surface, as measured through
the spectral density of the waves, has on the shape of the waves.
The second considers the asymmetry of real waves as compared to
simulated Gaussian waves, and also explores the effect that energy
may have on these differences.

The rest of this article is structured as follows. In section
\ref{sec2} a family of statistics for the two-sample problem for
functional data in introduced. Section \ref{sec3} gives a CLT for
these statistics and section \ref{sec4} gives application examples
to sea wave data.

\section{A family of statistics for the two-sample problem on
functional data} \label{sec2} Let $X_1(t),\dots,X_m(t)$ and
$Y_1(t),\dots,Y_n(t)$ be two functional data sets for which the null
hypothesis of equal distributions is to be evaluated. The $X_i$ and
$Y_j$ are assumed to live in a space of functions, $\X$, on the
interval $J$. Let $\tg_1,\dots,\tg_k$ be a finite set of functions
in $\X$. The $\tg_j$ might have been estimated using the $X$ and $Y$
samples. For fixed $g\in\X$, consider the dot product functional on
$\X$ defined by \[ G_g(x)=\int_J x(t)g(t)\mbox{d}t.
\]
Let $\lx$ and $\ly$ denote, respectively, the probability laws that
produce the $X$ and $Y$ samples. Then, the corresponding expected
values of $G_g(X)$ and $G_g(Y)$ are
 \begin{small}
\begin{equation}\label{l1} \lx(G_g)=\esp G_g(X)=\esp \int_J
X(t)g(t)\mbox{d}t\>\> \mbox{and} \>\> \ly(G_g)=\esp G_g(Y)=\esp
\int_J Y(t)g(t)\mbox{d}t
 \end{equation}
 \end{small}
where $X$ and $Y$ are random functions with distributions $\lx$ and
$\ly$, respectively. For each $g$, define the empirical processes
with respect to each sample by
\begin{small}
\beq\label{l2} \nu_X(G_g)=\frac{1}{\sqrt{m}}\left(\sum_{i\leq
m}(G_g(X_i)-\esp G_g(X))\right)\quad \mbox{and} \>\>
\nu_Y(G_g)=\frac{1}{\sqrt{n}}\left(\sum_{j\leq n}(G_g(Y_j)-\esp
G_g(Y))\right) \enq
\end{small}
For each $k$-tuple $\mathbf{g}=(g_1,\dots,g_k)$ of functions in $\X$
consider the empirical process vectors
 \begin{small}
 \beq\label{l3}
\nu_X(G(\mathbf{g}))=(\nu_X(G_{g_1}),\dots,\nu_X(G_{g_k}))^t\quad
\mbox{and} \quad
\nu_Y(G(\mathbf{g}))=(\nu_Y(G_{g_1}),\dots,\nu_Y(G_{g_k}))^t \enq
 \end{small}
Assume that for the vector $\mathbf{g}$ considered, the covariance
matrices of $\nu_X(G(\mathbf{g}))$ and $\nu_Y(G(\mathbf{g}))$, say
$C({X,\vg})$ and $C({Y,\vg})$, exist. For a given $\vg$, under the
null hypothesis, the matrices $C({X,\vg})$ and $C({Y,\vg})$
coincide. Write $C({\vg})$ for their common value. The class of
statistics that will be considered here are of the form:
\begin{equation}\label{l4}
 \begin{split}
Q_n&=\eta(m,n)^t\left(\tilde{C}({\tvg})\right)^{-1}\eta(m,n), \>\>\mbox{with}\\
\eta(m,n)&=\alpha(m,n)\nu_X(G({\tvg}))-\beta(m,n)\nu_Y(G({\tvg})),
 \end{split}
\end{equation}
where $\tvg=(\tg_1,\dots,\tg_k)$ is the (random) vector of functions
mentioned above and $\tilde{C}({\tvg})$ is a natural estimator of
the common covariance matrix for the empirical process vectors of
(\ref{l3}) evaluated on the functions of $\tvg$. The numbers
$\alpha(m,n)$ and $\beta(m,n)$ are chosen in such a way that the
expected values $\esp G({\tg_{j}})(X)$ and $\esp G({\tg_{j}})(Y)$
that appear in (\ref{l2}), cancel out (under the null hypothesis) in
the formula for $\eta(m,n)$, making unnecessary the estimation of
means. The rationale for considering this type of statistics is, we
believe, a natural one: Under the null hypothesis, the vectors
$\nu_X(G({\tvg}))/\sqrt{m}$ and $\nu_Y(G({\tvg}))/\sqrt{n}$ will
converge to the same limit (zero) as the sample sizes increase,
causing the quadratic form, $Q_n$, to be bounded in probability (it
will actually converge in distribution to a chi-square limit). Under
the alternative, for properly chosen functions $\tg_j$, there will
be no cancelation of the means, the norm of $\eta(m,n)$ will diverge
and, therefore, $Q_n$ will go to infinity.

A particular case of the statistic $Q_n$ is  the second method
proposed in  \citet[Ch. 5, p. 67]{hk}, where the functions in $\tvg$
are a subset of the principal components for the joint sample,
although the presentation of the statistic, the assumptions made
and, particularly, the methods of proof of properties differ from
those in the present article.

\section{A Central Limit Theorem for the statistics proposed}\label{sec3}
Note first that, by choosing \beq\label{l5}
\alpha=\alpha(m,n)=\frac{\sqrt{n+m}}{\sqrt{m}}\mbox{ and }
\beta=\beta(m,n)=\frac{\sqrt{n+m}}{\sqrt{n}} \enq the formula in the
second line of (\ref{l4}) reduces to
\[
\eta=\eta(m,n)=\frac{\sqrt{n+m}}{m}\sum_{i\leq
m}G_g(X_i)-\frac{\sqrt{n+m}}{n}\sum_{j\leq n}G_g(Y_j)
\]
making it unnecessary to compute (or estimate) the expectations for
the calculation of $\eta$. From here on, we will drop the subscripts
and write $\alpha$, $\beta$ and $\eta$, without specifying the
sample sizes, unless necessary. The functionals $G_g$ are defined on
the same underlying probability space of the $X$ and $Y$ functions
on which they are applied.

For the reader's convenience, we now recall the definitions of
``covering number'' and ``metric entropy''. Let $\F\subset L^p(Q)$,
for $p=1$ or 2, and a probability measure $Q$ on a probability
space. For $\eps>0$, the $\eps$-covering number of $\F$ with respect
to $Q$, $N_p(\eps,\F,Q)$, is the minimum natural $m$ such that there
exist functions $g_1,g_2,\dots,g_m\in L^p(Q)$ satisfying that, for
every $f\in \F$, there is a $j\in \{1,\dots,m\}$ such that
$\|f-g_j\|_{p,Q}<\eps$ where $\|\cdot\|_{p,Q}$ is the norm of
$L^p(Q)$. $H_p(\eps,\F,Q)=\log N_p(\eps,\F,Q)$ is called the metric
entropy of $\F$. For details on metric entropy and related notions
the reader can see \cite{dudc}, \cite{poll82}, \cite{poll},
\cite{vdv} or \cite{vdvw}.

We have the following proposition.

\begin{prop}\label{prop1}
Assume that the functions $g$ used to define $G_g$ are taken from a
class $\G\subset \X\subset L^2(J)$ (it will be convenient to assume
that $\G$ is included in $\X$). Assume as well the following: \neno
(i) There is a real valued function $F$ on $J$, such that, for all
$g\in\G$, and all $t\in J$, $|g(t)|<F(t)$ and $\|F\|^2_{2,J}=\int_J
F^2(t)\mbox{d}t< \infty$. \neno (ii) The random functions $X(t)$
satisfy  $\|X\|^2_{2,J}\leq M$, for some positive constant $M$.\neno
(iii) The collection $\G$ satisfies Pollard's entropy condition with
respect to Lebesgue measure:\beq\label{e6} \int_0^1\sqrt{\log
N_2(\epsilon\|F\|,\G,\lambda)}\mbox{d}\eps<\infty, \enq where
$\lambda$ is Lebesgue measure on $J$. Then, the class of functionals
\[ \H=\{G_g: g\in\G  \}
\]
is bounded by a constant $C$: for all $g\in \G$ and $X\sim\lx$,
$|G_g(X)|\leq C$. Furthermore, the class $\H$ satisfies Pollard's
uniform entropy condition:\beq\label{e7} \int_0^1\sqrt{\log
N_2(C\epsilon,\H)}\mbox{d}\eps<\infty \enq where
$N_2(C\epsilon,\H)=\sup_{{\cal L}^*}N_2(C\epsilon,\H,{\cal L}^*)$ is
a supremum over all probability measures ${\cal L}^*$ on the set of
functions where the $X(\cdot)$ live.
\end{prop}
\emph{Proof:} For each random function $X$ on $J$, and $G_g\in\H$,
by the Cauchy-Schwarz inequality, we have
 \beq\label{e7a}
 |G_g(X)|\leq\sqrt{\int
 X^2(t)\mbox{d}t\int g^2(t)\mbox{d}t}\leq\sqrt{M\int
 F^2(t)\mbox{d}t}.
 \enq
by hypothesis. Next, let $g^*_1,g^*_2,\dots,g^*_m$ be a minimal set
of functions such that, for every $g\in\G$, there exists $j\leq m$
for which $\|g-g^*_j\|_{2,J}\leq\eps$. Let ${\cal L}^*$ be a
probability measure on $\X$. Then, \beq\label{e8} {\cal
L}^*(G_g-G_{g^*_j})^2={\cal L}^*\left(\int X(t)(g-g^*_j)(t)\mbox{d}t
\right)^2\leq M\epsilon^2 \enq again by the Cauchy-Schwarz
inequality, and independently of the particular ${\cal L}^*$. It
follows that, for an appropriate choice of a positive constant
$\gamma$, \[ N_2(C\epsilon,\H)\leq N_2(\epsilon\gamma,\G,\lambda)
\]
and the result follows.

Today, there exist many ways of establishing upper bounds for the
metric entropy $\log N_2(\epsilon\|F\|,\G,\lambda)$ in Proposition
1. For instance, the process of ``registering'' the functional data
typically involves some degree of smoothing. When this is the case,
the functions to be analyzed and compared, that is, the functions in
$\X$, will be functions of bounded variation. Now, if $\G$ is a
class of functions of variation bounded by a fixed constant $D>0$,
then $\log N_2(\epsilon\|F\|,\G,\lambda)\leq K\epsilon^{-1}$, for
some positive constant $K$ (see Section 3 in \cite{vdv2}) and this
is enough for condition (\ref{e6}) to hold. On the other hand, in
our first example in Section 4.1, the functions in $\G$ are
indicators of intervals, which form a VC-subgraph class of functions
and, therefore, satisfy condition (\ref{e6}) comfortably (see
\cite{dudc} for the details). Proposition 1 tells us that these
metric entropy bounds will be inherited by the dot product
functional class, $\H$, a very convenient fact.

As for the distribution of $Q_n$ in (\ref{l4}), we have the
following:
\begin{prop}\label{prop2}
To the assumptions of Proposition 1 add the following: The functions
in the vector $\tvg=(\tg_1,\dots,\tg_k)$, appearing in the
definition of $Q_n$, converge in probability, in $L^2(J)$, to
limiting functions $(g_{1,\infty},\dots,g_{k,\infty})$ such that,
the covariance matrix of the limiting dot product functional vector,
$G(\mathbf{g}_\infty)(X)=(G_{g_{1,\infty}}(X),\dots,G_{g_{k,\infty}}(X))$,
say $C({X,\vg_\infty})$, exists and is not singular. From the
$X$-sample, assume that this covariance matrix is estimated as the
sample covariance, $\tilde{C}({X,\tvg})$ of the vectors
$(G_{\tg_{1}}(X_i),\dots,$ $G_{\tg_{k}}(X_i))$ for $i\leq m$ and
likewise, from the $Y$-sample, as $\tilde{C}({Y,\tvg})$. Suppose we
use as estimator of the covariance matrix $\tilde{C}({\tvg})$ in
(\ref{l4}), the appropriate multiple of the pooled covariance
matrix: \[
\tilde{C}({\tvg})=\frac{\alpha^2+\beta^2}{m+n-2}\big((m-1)\tilde{C}({X,\tvg})+(n-1)\tilde{C}({Y,\tvg})\big).
\]
Then, $Q_n$ converges in distribution to a chi-square variable with
$k$ degrees of freedom.
\end{prop}
\emph{Proof:} Under the null hypothesis of equality of
distributions, the matrices $C({X,\vg_\infty})$ and
$C({Y,\vg_\infty})$ are the same. We are writing $\mathbf{g}_\infty$
for the vector of the $g_{j,\infty}$, $j\leq k$ and
$G({\mathbf{g}_\infty})$ for the corresponding vector of dot product
functionals. Now, by Pollard's uniform Entropy Condition that holds
for $\H$, the Donsker property holds for the dot product class $\H$.
This means that the empirical processes $\nu_X(G(\mathbf{g}))$ and
$\nu_Y(G(\mathbf{g}))$, both indexed in $\G$, converge uniformly to
a limiting Gaussian process and, by Dudley's asymptotic
equicontinuity condition and the assumed convergence of the
functions in the vector $\tvg$,
 \[
 \nu_X(G({\tvg}))\cip \nu_X(G(\mathbf{g}_\infty)) \mbox{ and, likewise, }
 \nu_Y(G({\tvg}))\cip  \nu_Y(G(\mathbf{g}_\infty)).
 \]
Since the processes $\nu_X(G({\tvg}))$ and $\nu_Y(G({\tvg}))$ are
independent, the quadratic form $Q_n^*$, computed with the same
formula of $Q_n$, but using $C({X,\vg_\infty})$ as covariance
matrix, will have, by the Continuous Mapping theorem, the chi-square
distribution of the statement. Thus, by Slutzky's theorem, it only
remains to show that $\tilde{C}({X,\tvg})$ converges pointwise, in
probability, to $C({X,\vg_\infty})$. But using inequalities
(\ref{e7a}) and (\ref{e8}), it is easy to see that the covariance
matrix $C(X,\tvg)$ is a continuous function of the vector $\tvg$,
with respect to the norm of $L^2(J)$. Thus, by the triangle
inequality, it suffices to have a uniform law of large numbers for
the class \beq\label{e9} {\cal H}^{(2)}=\{G_g\,G_f:\> g,f\in{\cal
G}\} \enq and for the class ${\cal H}$ as well.  Now, let $\cls$ be
a probability law on $\X$ and $g,g',f,f'$ functions in $\G$. Then,
using Proposition \ref{prop1}, we get \beqa
\cls|G_g\,G_f-G_{g'}\,G_{f'}|&\leq &\cls(|G_g-G_{g'}||G_{f'}|)+\cls(|G_f-G_{f'}||G_{g}|)\nonumber\\
&\leq&C(\cls(|G_g-G_{g'}|)+\cls(|G_f-G_{f'}|)),\nonumber \enqa for
the constant $C$ in that Proposition. It follows that, \[
N_1(\epsilon,{\cal H}^{(2)},{\cal L}^*)\leq
N^2_1(\frac{\epsilon}{2C},{\cal H},{\cal L}^*)\leq
N^2_2(\frac{\epsilon}{2C},{\cal H},{\cal L}^*),
\]
and since the covering number $N_2({\epsilon}/{2C},{\cal H})$
satisfies Pollard's uniform entropy condition (\ref{e7}), the same
will hold for  $\sup_{\cls}N_1(\epsilon,{\cal H}^{(2)},{\cal L}^*)$
(squaring the covering number does not affect the entropy
condition), and this is more than enough for a Uniform Law of Large
Numbers for ${\cal H}^{(2)}$. The argument for ${\cal H}$ is simpler
and omitted, and the proof of Proposition \ref{prop2} is complete.

\section{Performance evaluation on examples}\label{sec4}

This section describes the application of the methodology presented
in Sections \ref{sec2} and \ref{sec3} on three problems from the
field of Oceanography. The first application is related to the
comparison of spectral densities, while the other two examples are
related to the analysis of the shape of waves.

\subsection{Comparison of spectral densities}\label{sec4-1}

As was mentioned in the Introduction, a frequent model for the sea
surface elevation at a fixed point is a centered stationary Gaussian
random process $X(t)$. The covariance $r(h) = \esp(X(t) X(t+h))$ of
this process has a spectral representation given by
$$
r(h) = \int e^{ih\omega} s(\omega)\, d\omega,
$$
where the function $s(\cdot)$ is known as the spectral density.
However, the stationarity hypothesis is not valid in the middle or
long term, and the use of stationary models is  limited in time,
depending on the weather conditions at the place of study. An
interesting and important problem is that of determining the
duration and characteristics of the stationary intervals for these
processes, and one possible point of view for this problem is the
analysis of the spectral densities estimated during short periods of
time.

Sea surface elevation data frequently come from
 moored buoys and the sampling
frequency is usually between 1 and 2 Hz. Data are stored in 20 or
30-minute intervals, which are considered to be short enough for the
stationarity assumption to hold, but long enough to have a good
estimation of the spectral density. Using this information, a
possible approach for the stationarity problem is to estimate the
spectral density for each time interval. Using the techniques
developed in the previous sections, one can compare, as we shall
see, the estimated spectral densities to determine whether they come
from the same distribution or not. If they do, and they are
contiguous in time, they correspond to a stationary period in the
wave data.

A simulation study was carried out in this context, in order
to compare spectral densities, using the Matlab
toolbox WAFO \citep{wafo} and spectra from the Torsethaugen
parametric family. This is a set of bimodal spectral densities of
frequent use in Oceanography, which account for the simultaneous
presence of wind-generated waves and swell, and was developed to
model spectra observed in the North-Sea. More details can be found
in \citet{Torset1} and \citet{Torset2}.

The parameters for the Torsethaugen family are the significant wave
height $H_s$ and the spectral peak period $T_p$. The significant
wave height is a standard measure of sea severity and is defined as
$H_s=4\sigma$, where $\sigma^2$, the variance of the process, is the
integral of the spectral density $s$:
\[
\sigma^2=\int s(\omega)\dife\omega.
\]
The spectral peak period is the inverse of the modal (peak)
frequency of the spectral density.

\begin{figure}[th]\label{Esp1}
\centering
  \includegraphics[height=5cm]{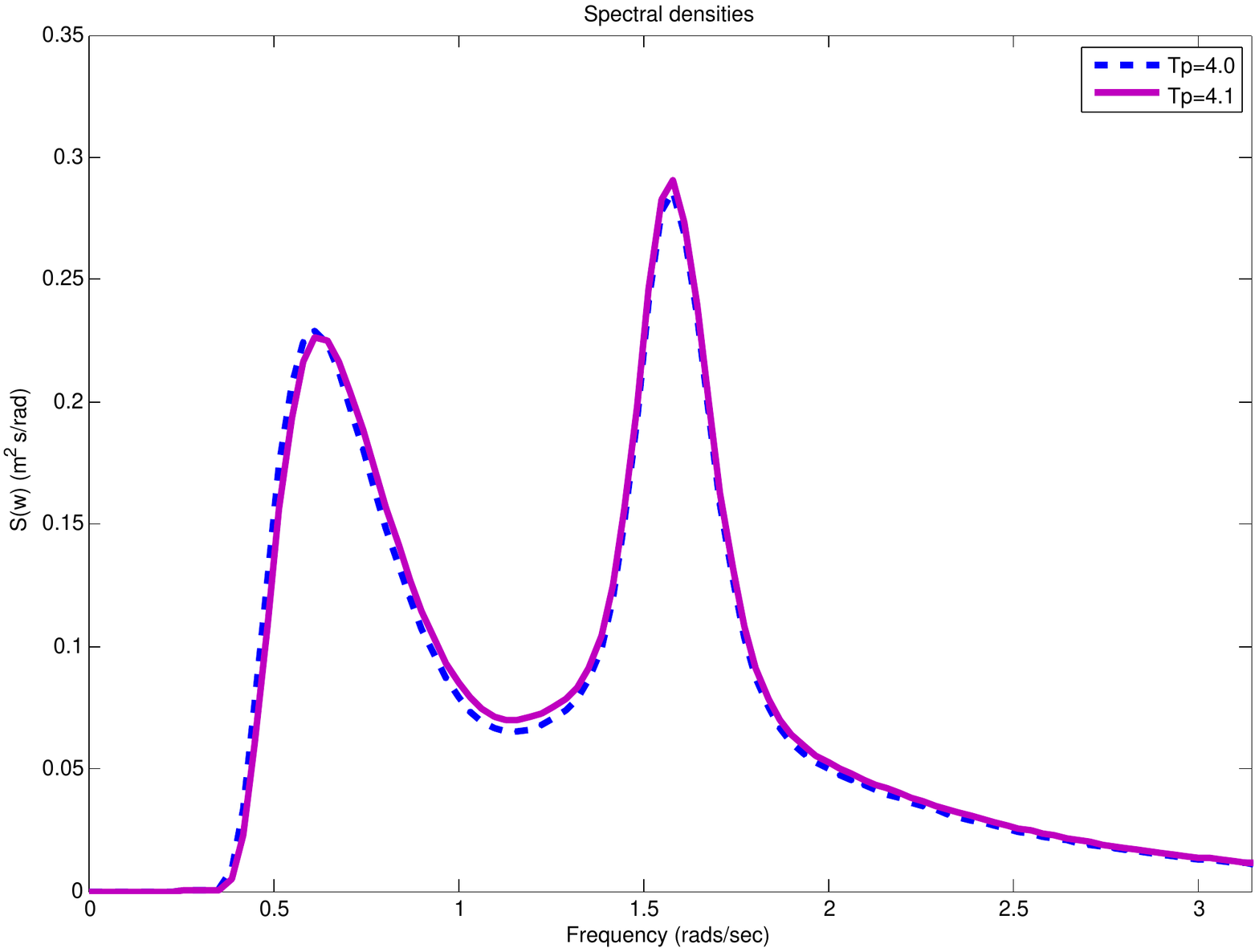}
  \includegraphics[height=5cm]{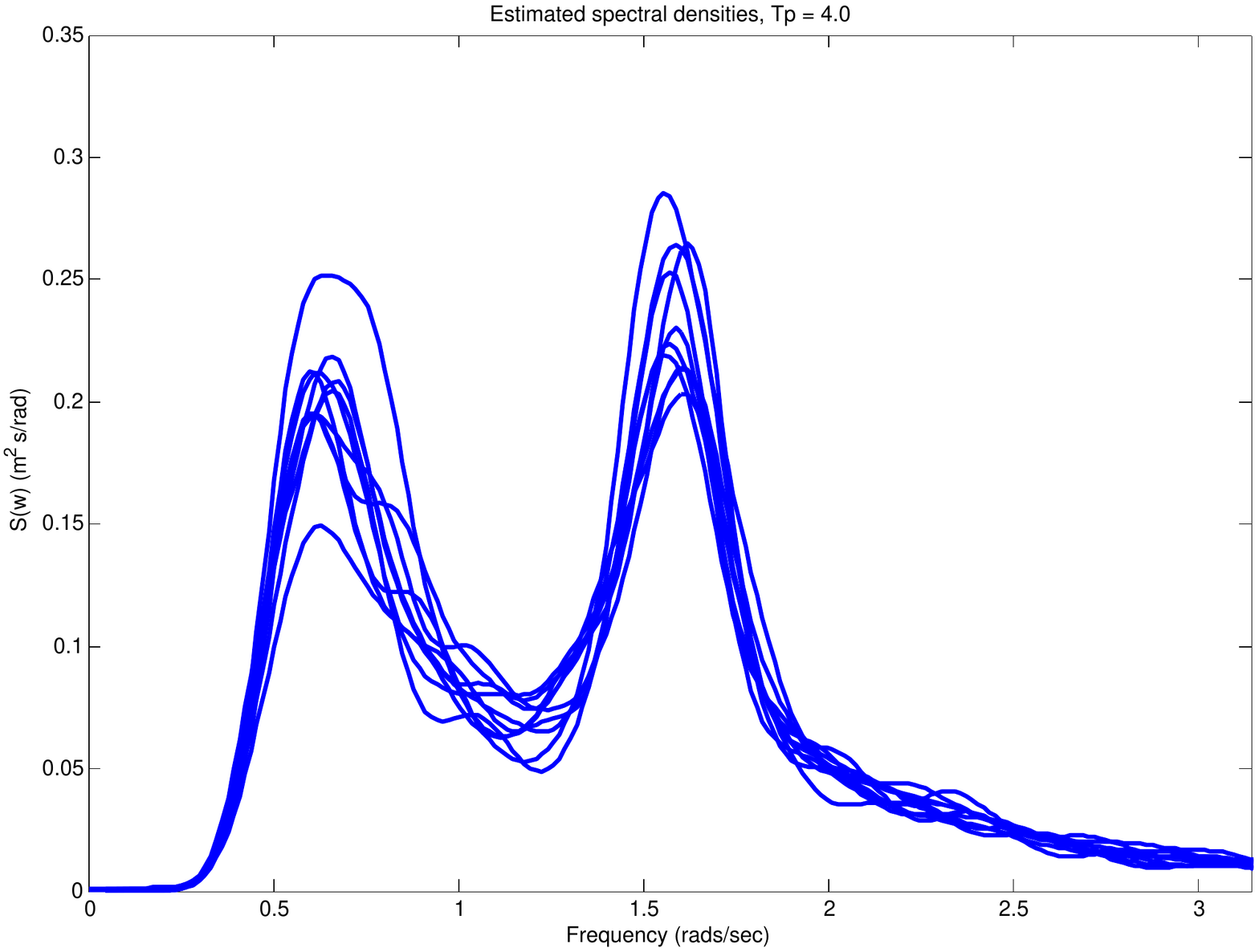}
  \includegraphics[height=5cm]{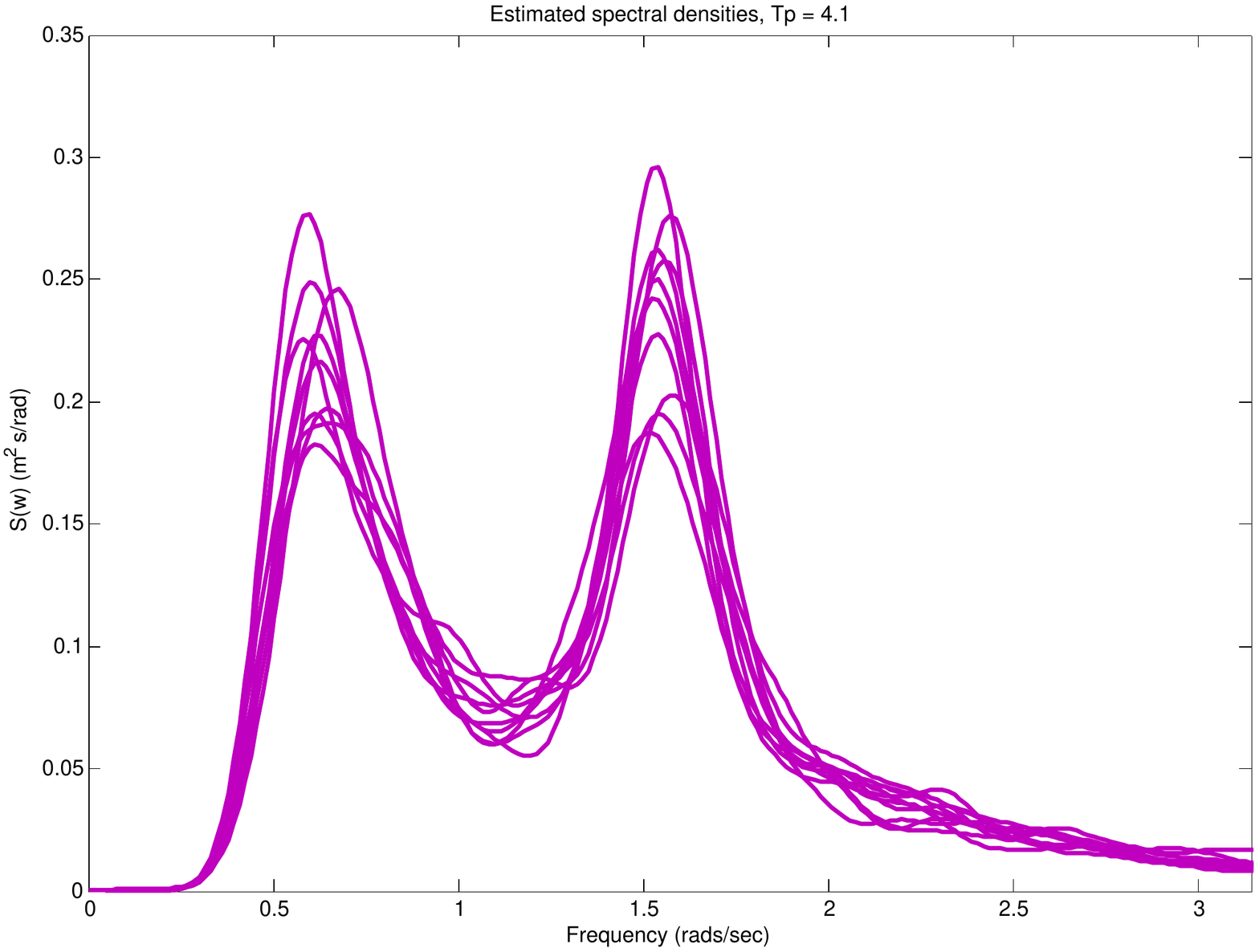}
  \caption{Torsethaugen spectra (left) and estimated
  spectral densities for $T_p=4.0$ (center) and $T_P=4.1$ (right).}
\end{figure}

The simulation scheme was as follows: Two spectral densities were
chosen from the parametric family. The parameters were set at
$H_s=2$ in both cases and $T_p= 4.0$ and $T_p=4.1$. Figure
\ref{Esp1} (left) shows the corresponding spectral densities. From
these densities and using the WAFO toolbox, stationary Gaussian
random (wave) processes lasting 30 minutes were simulated, with a
sampling frequency of 1.28  Hz., i.e., the time interval between two
consecutive points is 0.78125 seconds. These simulations correspond
to what would have
 been observed using a  moored buoy.

From each simulation, the spectral density was estimated using a
Parzen window with length 60. This was repeated 10 times for each of
the two original spectral densities, yielding two independent
samples of 10 functions each, which come from different populations.
Due to the random variations in the simulation and estimation
stages, these curves are similar in shape but present important
variations. Figures \ref{Esp1} (center) and (right)  show the two
samples. These densities were represented using a b-spline basis of
order 5 with 51 nodes in the interval $[0,\pi]$.

For testing whether the two samples come from the same distribution,
two versions of the $Q_n$ statistic were considered. For the first
one, the range of frequencies $[0,\pi]$ was divided into 8 intervals
and the indicator functions of these intervals play the role of the
$g$ functions. In this case, the score corresponding to a projection
along the direction of one of these indicator functions is
equivalent to integrating the energy for the range of frequencies
represented by the interval. For the second version, a b-spline
basis of order 5 with 7 interior nodes was used as $g$ functions.
The values obtained for $Q_n$ for these two versions of the
statistic were 39.105 and 120.18, respectively, with corresponding
$p$-values, with respect to the asymptotic distribution, of
$4.7\times 10^{-6}$ and 0.  Nevertheless, taking into account the
small sample sizes, asymptotic $p$-values cannot be considered valid
and we must resort to Monte Carlo $p$-values. For this purpose, we
use the fact that an algorithm is available for the generation of
independent samples with a given spectral density. From the
``original'' set of 20 estimated spectral densities (10 for each
parameter choice) an average spectral density was estimated, say
$s_{\mbox{avg}}$. Then, $s_{\mbox{avg}}$ was used to produce two
sets of 10 simulated stationary Gaussian random (wave) processes
lasting 30 minutes, using the WAFO toolbox, as described above. From
each 30 minute simulated wave process, the corresponding spectral
density was estimated, to produce two sets of 10 spectral densities
under the null hypothesis. On these two samples of spectral
densities the statistic $Q_n$ was computed. The simulation procedure
just described was repeated 10,000 times, using always the same
$s_{\mbox{avg}}$, and the 10,000 values of $Q_n$ produced were used
to estimate the $p$-value for the original value of the statistic.
The resulting Monte Carlo $p$-values were 0.0492 and 0.0398, for the
two schemes of $g$ functions considered, which shows that even for
the small sample sizes considered and for two similar Torsethaugen
spectral densities, the method proposed is able to produce some
evidence of difference.

Next, we performed the same simulation experiment, with larger
sample sizes, in order to assess speed of convergence to the
limiting distribution. In this case we considered a sample size of
140 estimated spectral densities for $T_p=4.0$ and 160 for
$T_p=4.1$. The $Q_n$ values for the sample were 507.03 for the
indicator basis and 517.16 for the b-spline basis, with $p$-values
equal to 0 in both cases. The Monte Carlo procedure is identical to
the one described above for the smaller sample sizes and, in this
case, from the Monte Carlo simulations, approximate quantiles were
estimated, along with the Monte Carlo $p$-values. The results,
regarding quantiles, for both versions of $Q_n$, are given in table
\ref{tab4-1}.

\begin{table}[ht]
\centering
\begin{tabular}{|c|c|c|c|c|c|}
\multicolumn{6}{l}{Spline basis}\\
 \hline Quantile   & 0.5 & 0.9 & 0.95 & 0.975 & 0.99\\
 \hline Asymptotic & 11.34  & 18.549 & 21.026 & 23.337 & 26.217\\
 \hline MC   & 11.857 & 19.451 & 22.242 & 24.846 & 27.445\\
 \hline Rel. error & -0.046 & -0.049 & -0.058 & -0.065 & -0.047\\
 \hline
\multicolumn{6}{l}{\ }\\
\multicolumn{6}{l}{Indicator basis}\\
 \hline Quantile   & 0.5 & 0.9 & 0.95 & 0.975 & 0.99\\
 \hline  Asymptotic & 7.344  & 13.362 & 15.507 & 17.535 & 20.09\\
 \hline  MC   & 7.443 & 13.819 & 16.064 & 18.202 & 20.892\\
 \hline  Rel. error & -0.013 & -0.034 & -0.036 & -0.038 & -0.040\\
 \hline
\end{tabular}
\caption{Finite sample and limiting quantiles for the spectral
density data.} \label{tab4-1}
\end{table}

The results in Table \ref{tab4-1} show that the relative errors,
between Monte Carlo and asymptotic quantiles, vary between 1.3 and
6.5\% and are always negative, indicating that the asymptotic values
 underestimate the true quantiles of the statistic in this case.
Thus, for this problem and the choices of functions $g$ made for
$Q_n$, we conclude that the convergence to the limit distribution of
the statistic is slow and it is advisable to calculate the necessary
$p$-values using a Monte Carlo procedure.

\subsection{Waves as functional data}\label{sec4-2}

The other two examples in this section concern the analysis of waves
as real functions. The raw data consists of sea surface elevation
measurements at a fixed point obtained from the U.S. Coastal Data
Information Program (CDIP) website. The data come from buoy 106
(51201 for the National Data Buoy Center), a station located at
Waimea Bay, Hawaii, at a sea depth of 200 meters. The surface
elevation was sampled at a frequency of 1.28 Hz, during 30-minute
intervals. A total of 430 intervals (8 days and 23 hours), between
January 1$^{\mbox{st}}$ and January 9$^{\mbox{th}}$, 2003, were
considered.

A wave is defined as the curve of surface elevation values between
two consecutive downcrossings of the mean sea level (see Figure
\ref{fig02}). For each 30-minute interval, the individual waves were
considered as functions. Since the time length of each individual
wave (the period) is different, all waves were registered to the
$[0, 1]$ interval by a  linear transformation of the time interval.
After registration, waves were initially represented using a
B-spline basis of order 6 with nodes at the data points that define
each wave. Then, these functions are represented using a common
basis, again B-splines of order 6, but with 61 equidistant nodes on
the interval $[0, 1]$, so that all waves have a representation in
terms of a common basis. The order of the splines guarantees that
the functions are smooth, having two continuous derivatives.

\begin{figure}[th]\label{fig2}
\centering
  \includegraphics[width=10cm]{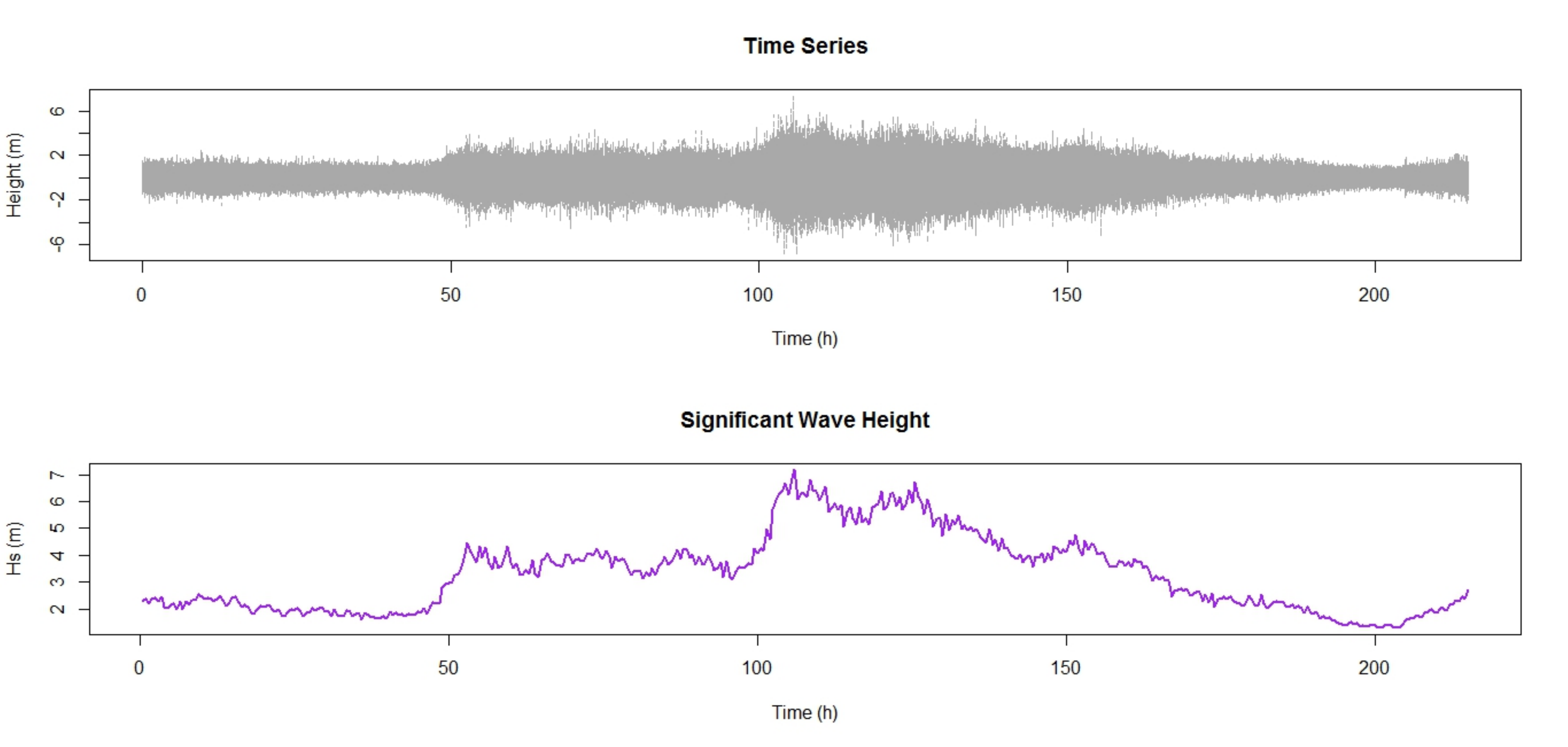}\\
  \caption{Wave height (top) and significant wave height (bottom) for Buoy 106.}\label{fig1}
\end{figure}

Spectral densities were estimated for each 30-minute interval using
the toolbox WAFO and  the values of $\sigma^2$ and $H_s$ were
obtained for each data interval. Figure \ref{fig1} shows both the
original sea surface elevation data and the evolution of $H_s$ in
time.

For the purpose of evaluation of the methodology proposed here, the
30-minute intervals were divided into four groups, $G1, G2, G3$ and
$G4$, according to the value of their significant wave height; the
groups correspond to values in the ranges $0-2$ m., $2-4$ m., $4-6$
m. and values over 6 meters for $G4$.

Within each energy group, two consecutive 30-minute intervals were
selected, and the waves corresponding to those intervals constitute
the data set for the group. The selected sets of waves are indicated
in Figure \ref{fig2} (left) and are denoted in what follows as
$H_s1$, $H_s2$, $H_s3$ and $H_s4$, with significant wave height
increasing with numbering. The number of waves in each one-hour set
are, respectively, 166, 171, 179 and 187. Figure \ref{fig2} (right)
shows the waves in the sets $H_s1$ to $H_s4$. It is clear that, in
terms of amplitude, the waves in these groups are different, with
the possible exception of $H_s3$ versus $H_s4$. We wish to quantify
these differences with the methodology proposed in the previous
section. We will compare, in the context of the two-sample problem,
$H_s1$ versus $H_s2$, $H_s2$ versus $H_s3$ and $H_s3$ versus $H_s4$.

\begin{figure}
 \centering
  \includegraphics[width=5cm]{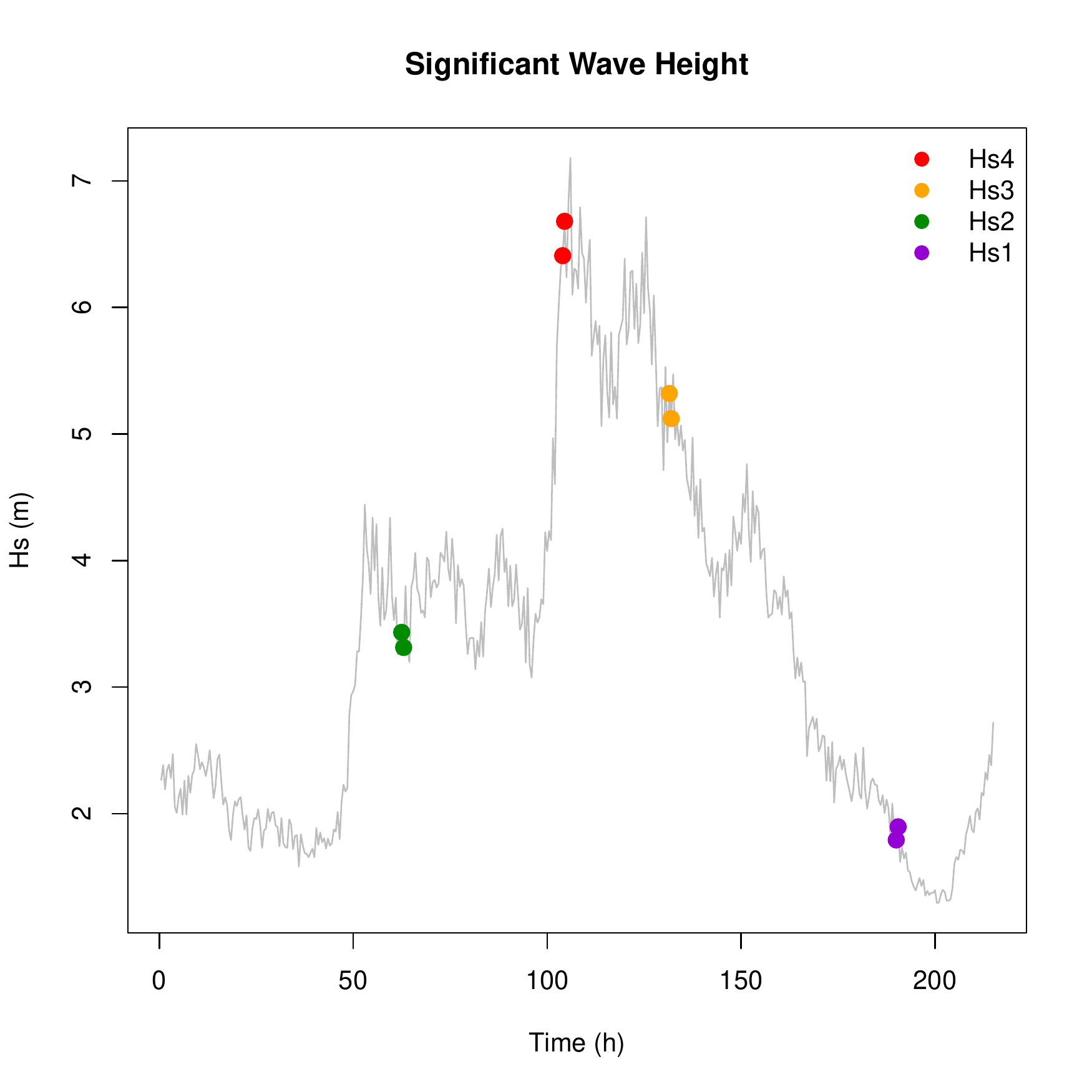} \quad
  \includegraphics[width=5cm]{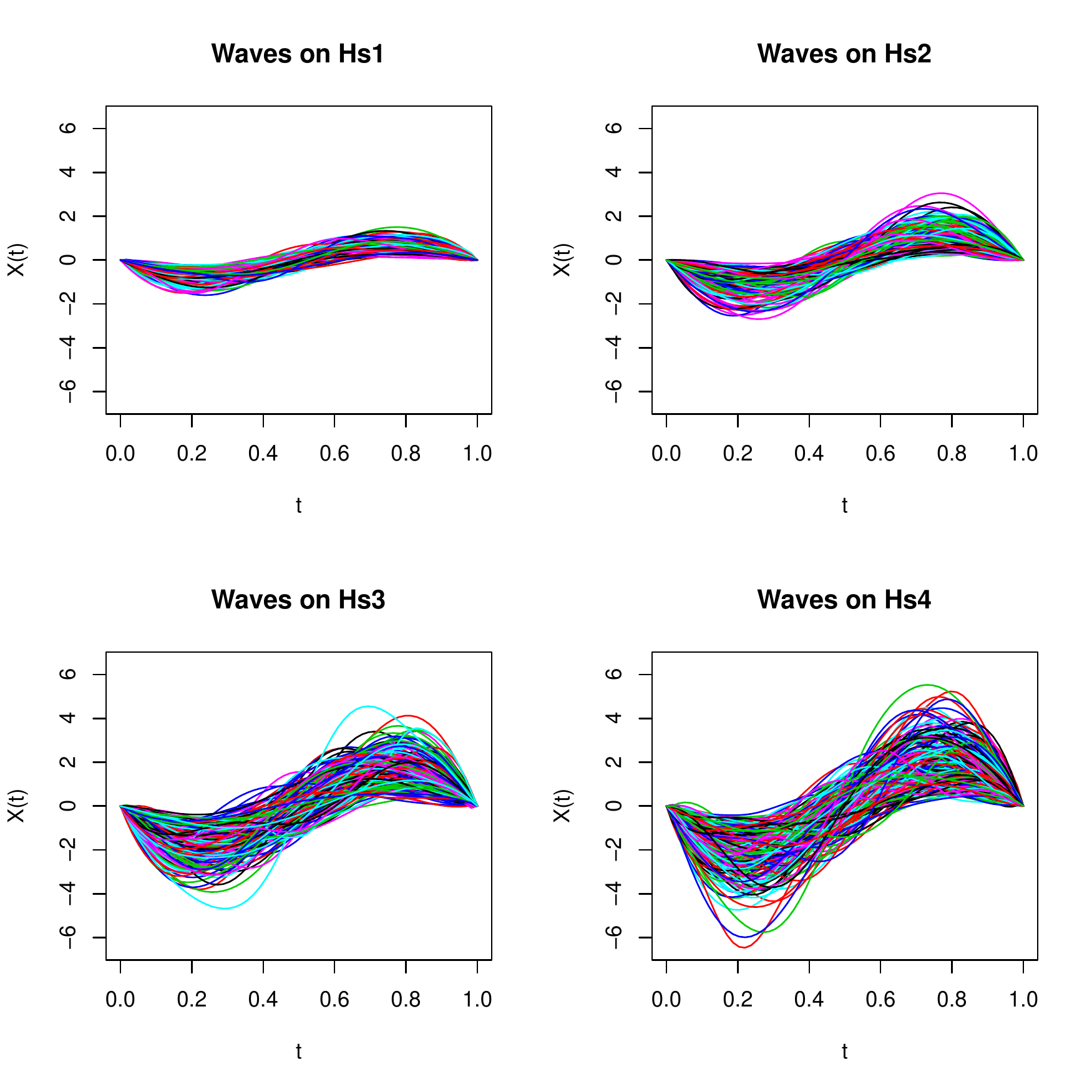}
  \caption{Position of selected intervals (left),
  waves in the selected time intervals (right).}\label{fig2-3}
\end{figure}

\subsection{Projection on odd and even trigonometric functions}\label{sec3-3}
\begin{figure}[t]
\centering
  \includegraphics[height=5cm]{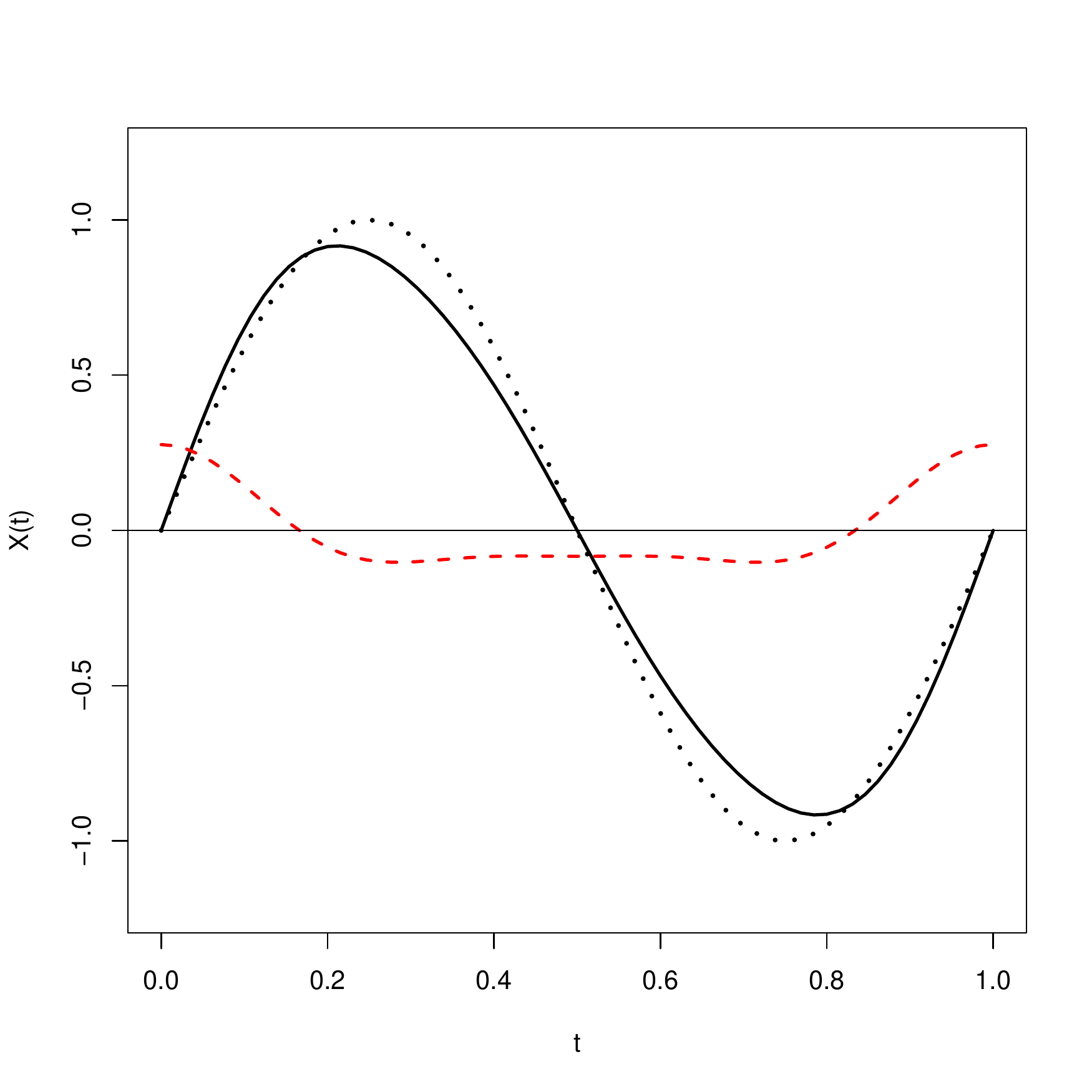} \quad
  \caption{$\tilde g_1$ (solid), $\tilde g_2$ (dashed) and sine function
  (dotted) for the $H_s1$ vs. $H_s2$ test.}
  \label{fig4-3}
\end{figure}

For the problem of comparing the different data sets described in
section \ref{sec4-2}, we apply the statistic $Q_n$, using two
functions in the vector $\tvg$, that will be certain projections of
the joint data set on linear combinations of the odd and even
trigonometric functions with coefficients determined from the joint
sample of functions. For each registered wave, $Z_i(t)$, in the
joint sample, we consider its $l$-th sine and cosine Fourier
coefficients, given by
 \beq\label{r08}
 a_{il}=\int_0^1 Z_i(t)\sin(2\pi l t)\dd t \qquad\mbox{ and }\qquad
 b_{il}=\int_0^1 Z_i(t)\cos(2\pi l t)\dd t.
 \enq
We compute these coefficients for $l\leq k=3$, since the
coefficients decrease very rapidly. For each $l\leq k$, we take the
averages of the absolute values of the $a_{il}$ and $b_{il}$ as
representatives of the relevance of the $l$-th term in the
expansion: \beq\label{r09}
\bar{a}_{l}=\frac{1}{N}\sum_{i=1}^{N}|a_{il}| \qquad\mbox{ and
}\qquad \bar{b}_{l}=\frac{1}{N}\sum_{i=1}^{N}|b_{il}|, \qquad\mbox{
for } l=1,2\mbox{ and }3, \enq where, as before, $N=m+n$ is the size
of the joint sample. Then, we take as our functions, $\tg_1$ and
$\tg_2$, the following
 \beq\label{r10}
 \tilde{g}_{1}=\sum_{l=1}^{3}\bar{a}_{l}\sin(2\pi l t)  \qquad\mbox{and}\qquad
 \tilde{g}_{2}=\sum_{l=1}^{3}\bar{b}_{l}\cos(2\pi l t).
 \enq
These functions are calculated for each pair of energy levels:
$H_s1$ versus $H_s2$, $H_s2$ versus $H_s3$ and $H_s3$ versus $H_s4$.
As a reference, Table \ref{tab3} shows the values of the
coefficients that define $\tg_1$ and $\tg_2$ in the case of $H_s1$
versus $H_s2$ and Figure \ref{fig4-3} shows the corresponding
$\tilde g_1$ and $\tilde g_2$ functions plus a sine function for
comparison purposes. Table \ref{tab4} shows the values and
$p$-values obtained when $Q_n$ is calculated for testing the
difference of distribution between the groups of waves of different
energy. Note that this time we evaluate the value obtained for $Q_n$
against the chi-square distribution with 2 degrees of freedom.

The numbers in Table \ref{tab4} reflect very strong evidence against
the null hypothesis in all cases, especially in the first two, as
could be expected from the waveforms in Figure \ref{fig2-3} (right),
and these results also say that  the projections on the
trigonometric functions are enough, through the $Q_n$ statistic, for
detecting the difference in energy between the samples.

\begin{table}[ht]
\centering
\begin{tabular}{|c|c|c|}\hline
j & $\bar{a}_{j}$  & $\bar{b}_{j}$  \\  \hline
1 & 0.916  &  0.151 \\
2  & 0.097    & 0.097 \\
3  &   0.024 &   0.029 \\
\hline
\end{tabular}
\caption{Coefficients defining $\tg_1$ and $\tg_2$ for the $H_s1$
vs. $H_s2$ test.} \label{tab3}
\end{table}

\begin{table}[ht]
\centering
\begin{tabular}{|c|c|c|} \hline
Pair of samples tested & $Q_n$ value & $p$-value \\  \hline
$H_s1$ versus $H_s2$ & 103.75  & 0 \\
$H_s2$ versus $H_s3$ & 107.37  & 0 \\
$H_s3$ versus $H_s4$ & 17.01 & 2.02$\times 10^{-4}$ \\
\hline
\end{tabular}
\caption{$Q_n$ values for projections on odd and even trigonometric
functions.}\label{tab4}
\end{table}

\begin{figure}[t]
 \centering
  \includegraphics[width=3.5cm]{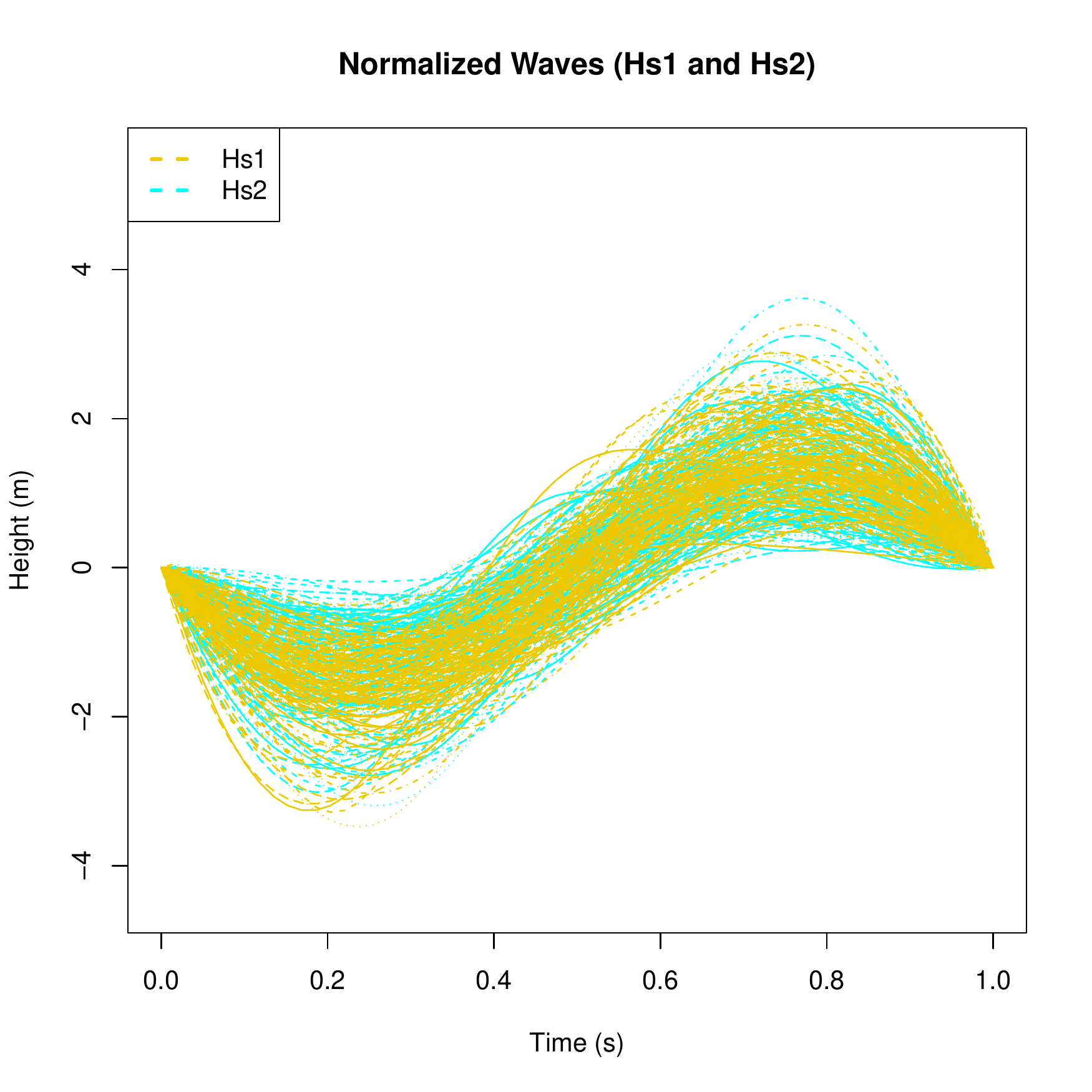} \quad
  \includegraphics[width=3.5cm]{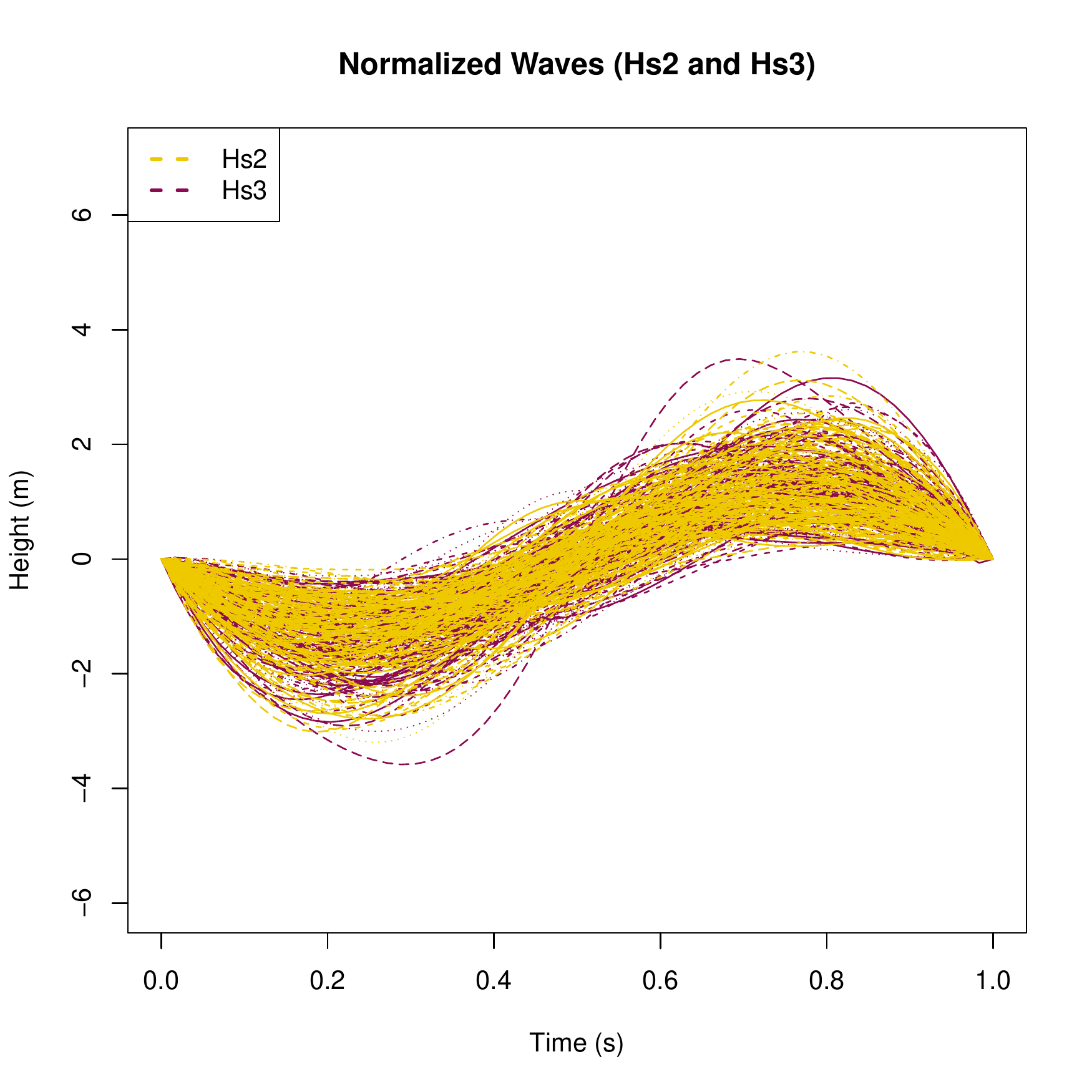} \quad
  \includegraphics[width=3.5cm]{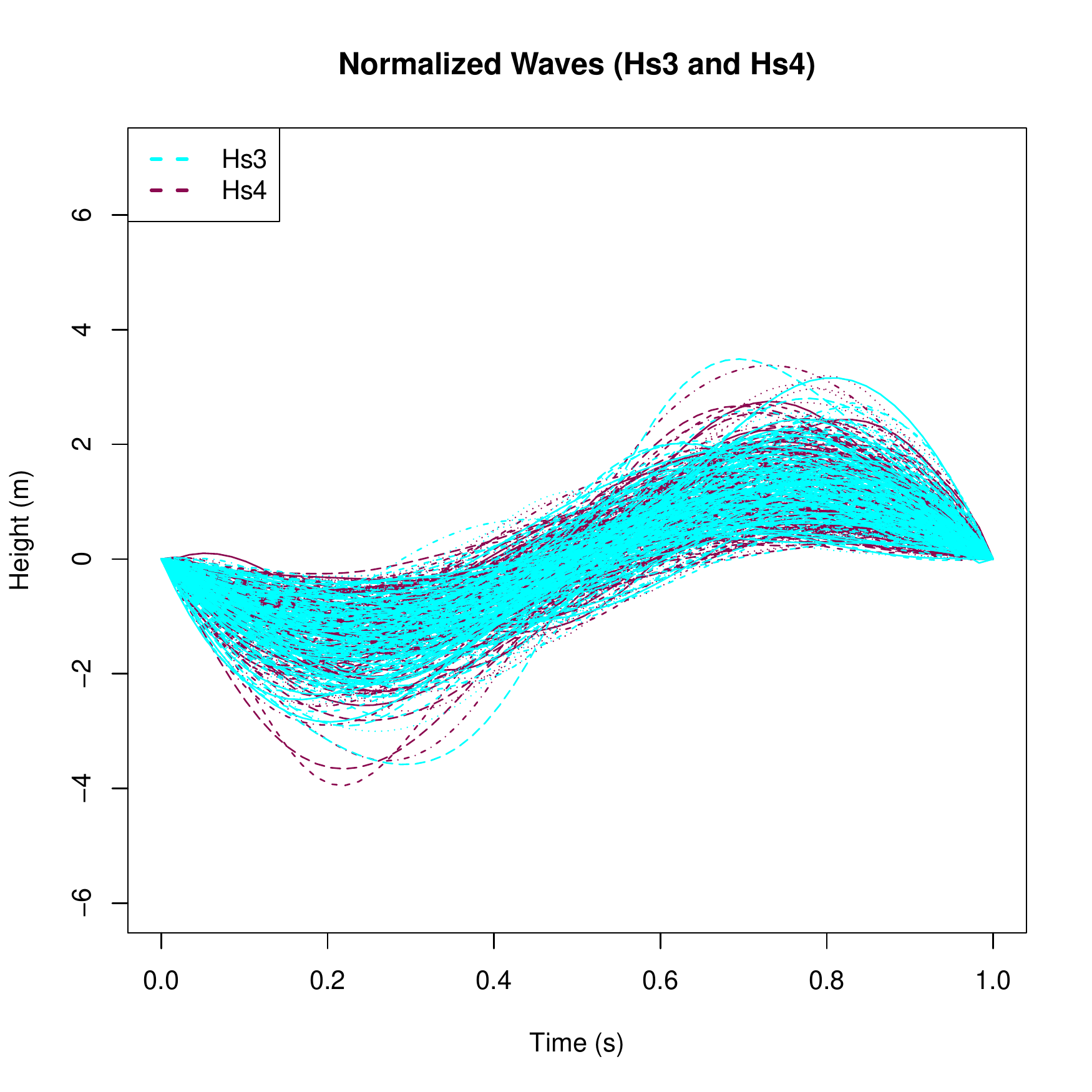}
  \caption{Waves in intervals $H_s1$ and $H_s2$ (left), $H_s1$ and
  $H_s3$ (center) and $H_s2$ and $H_s3$ (right).}\label{fig2-4}
\end{figure}

This result, however, is to be expected from the differences in
amplitude that can be observed in Figure \ref{fig2-3}. So a natural
question is whether the dissimilarities are only in amplitude, or
whether there are also differences in the shape of the waves due to
the variation in energy levels. To test if there are differences
between these samples other than in amplitude, the normalized waves
were considered, where the normalization was obtained dividing by
the standard deviation estimated for each one-hour interval. We
consider intervals $H_s1, H_s2$ and $H_s3$ and Figure \ref{fig2-4}
shows the registered waves for the three possible pairs of
normalized samples. The differences among them, if there are any,
are not so obvious now.

Using the same method as before, the three pairs of samples were
compared to test whether the curves come from the same distribution.
The results of these tests are given in table \ref{tab1b}, where the
$p$-values obtained using the asymptotic distribution and a
bootstrap procedure, described below, are included.

\begin{table}[ht]
\centering
\begin{tabular}{|c|c|c|c|} \hline
Pair of samples tested & $Q_n$ value & $p$-value (asymp.) & $p$-value (bootstrap) \\
\hline
$H_s1$ versus $H_s2$ & 17.15  & $1.88\times 10^{-4}$ & $4\times 10^{-4}$ \\
$H_s2$ versus $H_s3$ & 1.023  & 0.5997 & 0.5997 \\
$H_s3$ versus $H_s4$ & 2.258  & 0.323  & 0.333 \\
\hline
\end{tabular}
\caption{$Q_n$ and $p$-values based on principal components for the
normalized samples.} \label{tab1b}
\end{table}

\begin{figure}[t]
 \centering
  \includegraphics[width=7cm]{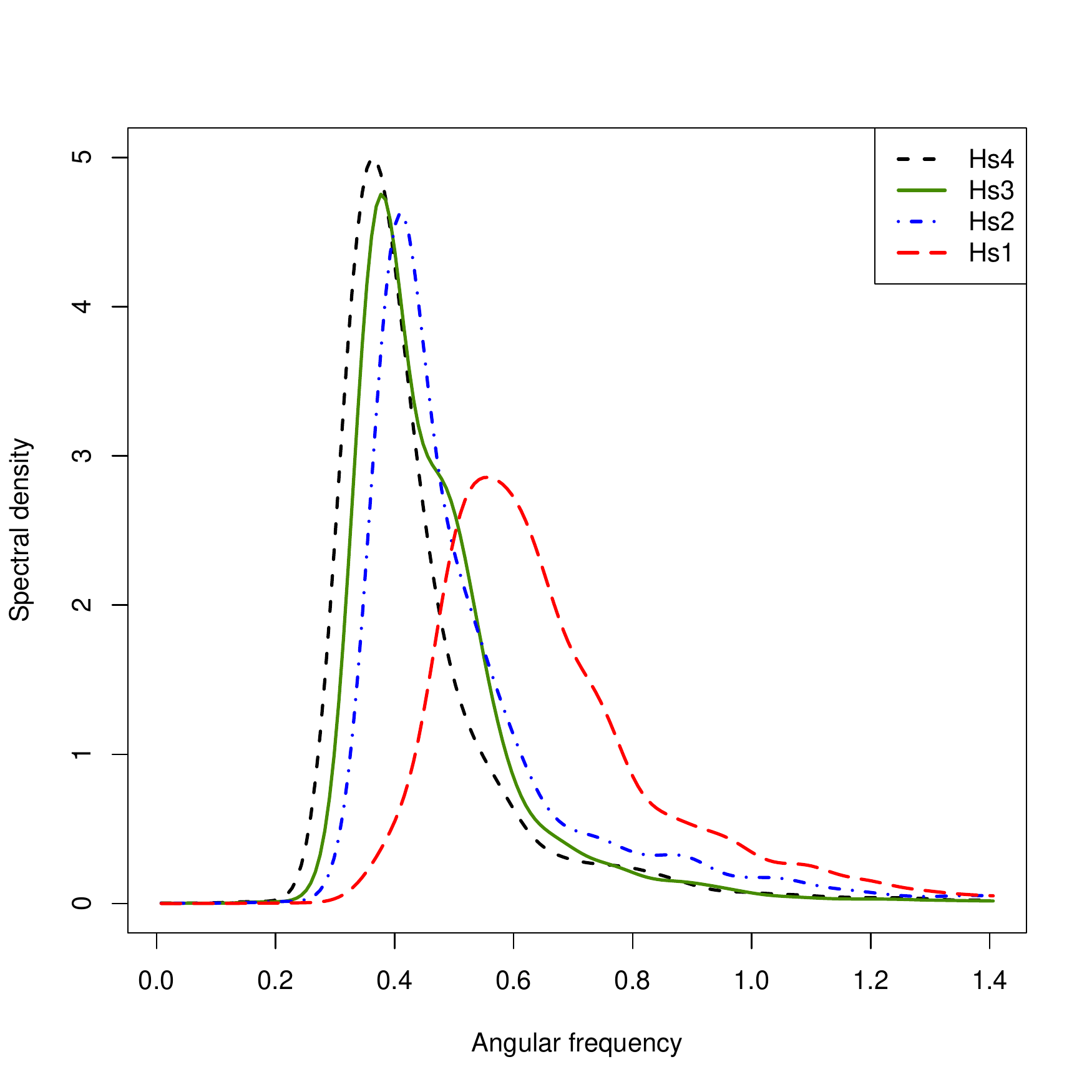}
  \caption{Spectral densities for intervals $H_s1, H_s2$ and $H_s3$.}\label{fig2-5}
\end{figure}

These values show that the differences are not so clear after
normalization, and point to the first interval $H_s1$ being
different from the other three, but there is no evidence of
differences between $H_s2$ and $H_s3$ or $H_s3$ and $H_s4$. Figure
\ref{fig2-5} shows the normalized spectra for these intervals,
and may help explain the results obtained, since the spectral
density of a time series sums up its oscillatory behavior. As can be
seen, the spectra for intervals $H_s2$, $H_s3$ and $H_s4$ are
similar in dominant frequency and dispersion while the spectrum for
$H_s1$ is clearly different in both aspects. It is important to
observe, however, that the process of registration of the individual
waves to a common interval losses the information about the period
of the wave, and hence also about frequency. Thus it seems likely
that it is the dispersion (and shape) of the spectral density rather
than its location in the frequency scale which accounts for the
differences observed in the three samples. Nevertheless, the
relationship between spectral densities and the shape of waves is
not clear and requires further exploration.

Next, we evaluate whether, in this example, the asymptotic
distribution as a reference is valid for the sample sizes
considered. Thus, our next experiment evaluates, through a bootstrap
procedure, the approximation to the null distribution in the present
context. The bootstrap method used here does not require the
estimation of spectral densities and, for this reason, is
computationally significantly less expensive than the Monte Carlo
procedure described before.

For this purpose, the 166 waves of data set $H_s1$ were used. The
data were randomly split into two sets of 106 and 60 waves,
respectively, and the $Q_n$ statistic was computed. This procedure
was repeated for 10,000 random selections of the two subsets, of 106
and 60 waves (from the same joint sample of 166), calculating $Q_n$
every time. From the 10,000 values, we obtain quantiles of $Q_n$
that correspond, approximately, to the null hypothesis and are
displayed in Table \ref{tab2}, were we have included, for comparison
purposes, the corresponding quantiles for the $\chi_2^2$
distribution.

\begin{table}[ht]
\centering
\begin{tabular}{|c|ccccc|} \hline
& \multicolumn{5}{|c|}{Probabilities} \\
             & .5    & .9    & .95   & .975  & .99 \\ \hline
 asymptotic  & 1.386 & 4.605 & 5.992 & 7.378 & 9.21 \\  \hline
 bootstrap   & 1.365 & 4.662 & 6.147 & 7.454 & 9.169 \\ \hline
 relative error & 0.0157 & -0.0124 & -0.0259 & -0.0103 & 0.0045 \\
 \hline
\end{tabular}
\caption{Finite sample and limiting quantiles of $Q_n$ for $H_s1$
data.} \label{tab2}
\end{table}
The good agreement between finite sample and limiting quantiles in
Table \ref{tab2}, suggest that for sample sizes above a hundred for
both samples, the proposed statistic can be confidently used for the
type of data considered in this example. This form of bootstrap was
used to produce the ``bootstrap'' quantiles in Table \ref{tab1b}  by
bootstrapping from the joint dataset in each case.

\subsection{Asymmetry of real waves}
One of the advantages of the set of statistics proposed in this work
is its flexibility. In this section we show how it is possible to
construct statistics suitable for the assessment of
symmetry in samples of functions. In \cite{goros}, sets of
registered storm waves were evaluated for asymmetry and also
compared, by means of a conditional permutation test, to sets of
waves generated from the Gaussian model with parameters estimated
from the data. The Gaussian model, as described for instance in
\cite{ochi}, is a standard stochastic model for sea waves. Still, it
was pointed out in the introduction that real waves differ from
those produced by the model, in that real waves present more
asymmetry than the model would allow, having shallower troughs and
more peaked crests, and this difference may be more marked at higher
energy levels.

We will now use the proposed statistic $Q_n$ to test for the null
hypothesis that registered real waves and waves produced by the
Gaussian model have the same distribution against the alternative
that real waves show more asymmetry. For this purpose, we take a set
of waves corresponding to two consecutive 30 minute period in each
of the four energy levels considered in Section 4.2. For this
analysis, in the registration process of the waves an added
restriction was that the upcrossing of the mean level occurs at 0.5.
The precise description of the registration procedure can be found
in \cite{goros}. From each dataset, the spectral density is
estimated and from it, a set of simulated waves is produced, using
the WAFO toolbox of the MATLAB language. For each energy level, from
the combined sample (real and simulated waves), we compute, for
$l\leq 3$ and $i\leq N$, the coefficients $a_{il}$, the average
$\bar{a}_{l}$ and the function $\tilde{g}_{1}$ of (\ref{r08}),
(\ref{r09}) and (\ref{r10}). The idea is the following: The
suspected asymmetry of real waves consists on the waves being
less deep, in terms of amplitude, during the first half cycle, than
the amount they rise above sea level on the second half cycle. This
type of asymmetry should show in the dot product against a relevant
odd function. Thus, we estimate a representative odd function
$\tilde{g}_{1}$ and apply the statistic $Q_n$ with that function
alone. If the alternative hypothesis holds, we expect that, for the
real waves, the dot product with $\tilde{g}_{1}$ will have a
positive mean value, while it will tend to zero for the simulated
waves.

\begin{figure}
\centering
  \includegraphics[width=5cm,height=4cm]{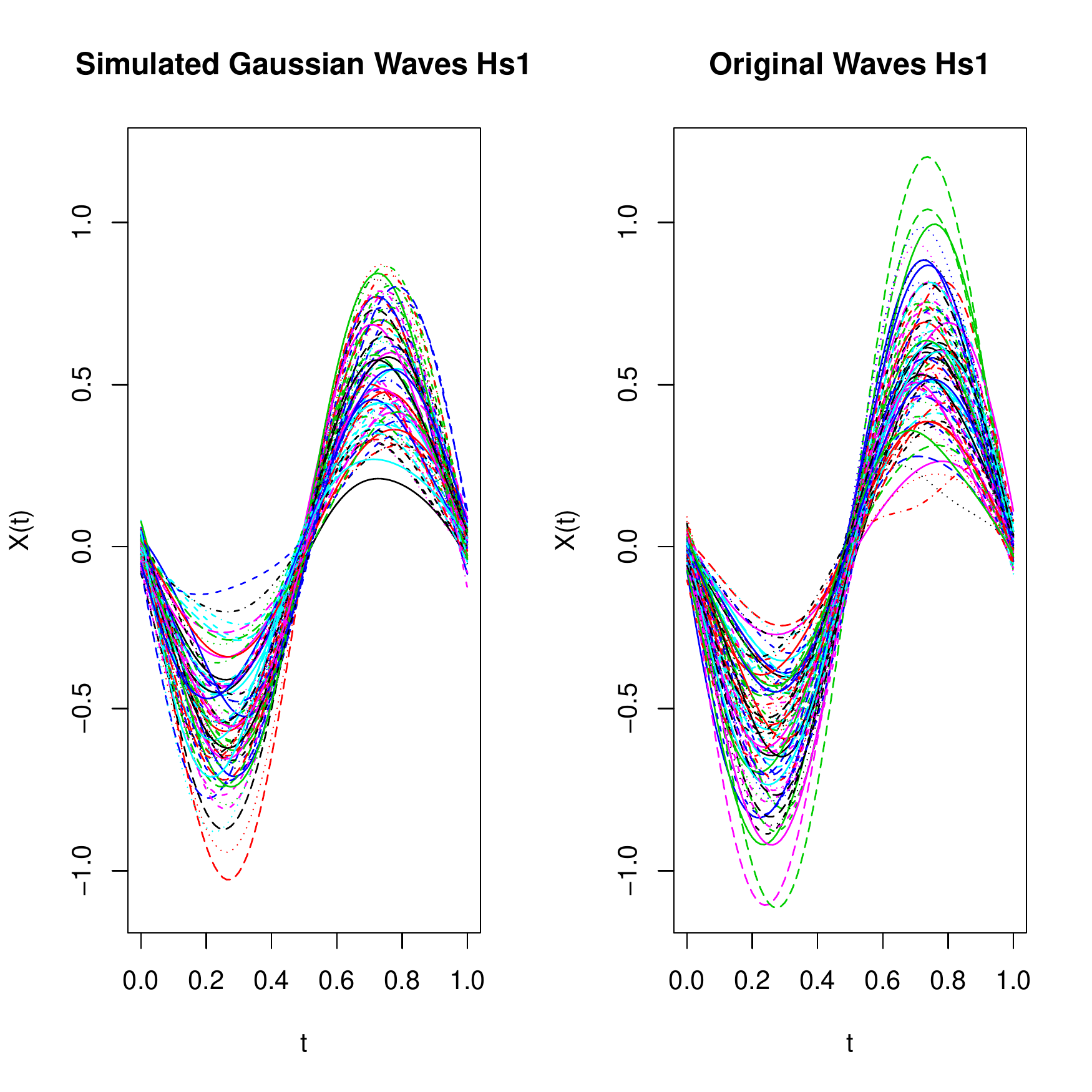} \quad
  \includegraphics[width=5cm,height=4cm]{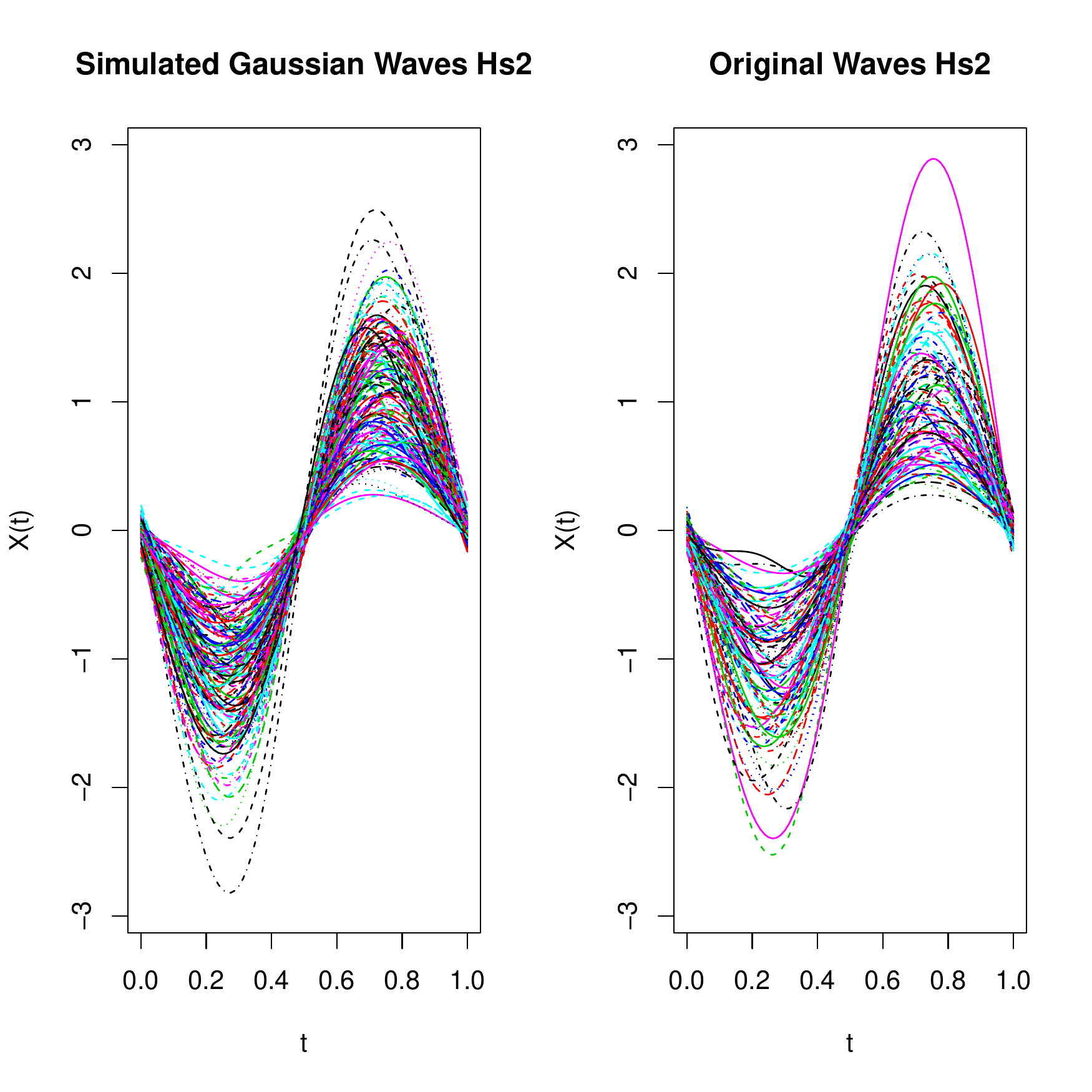} \\
  \includegraphics[width=5cm,height=4cm]{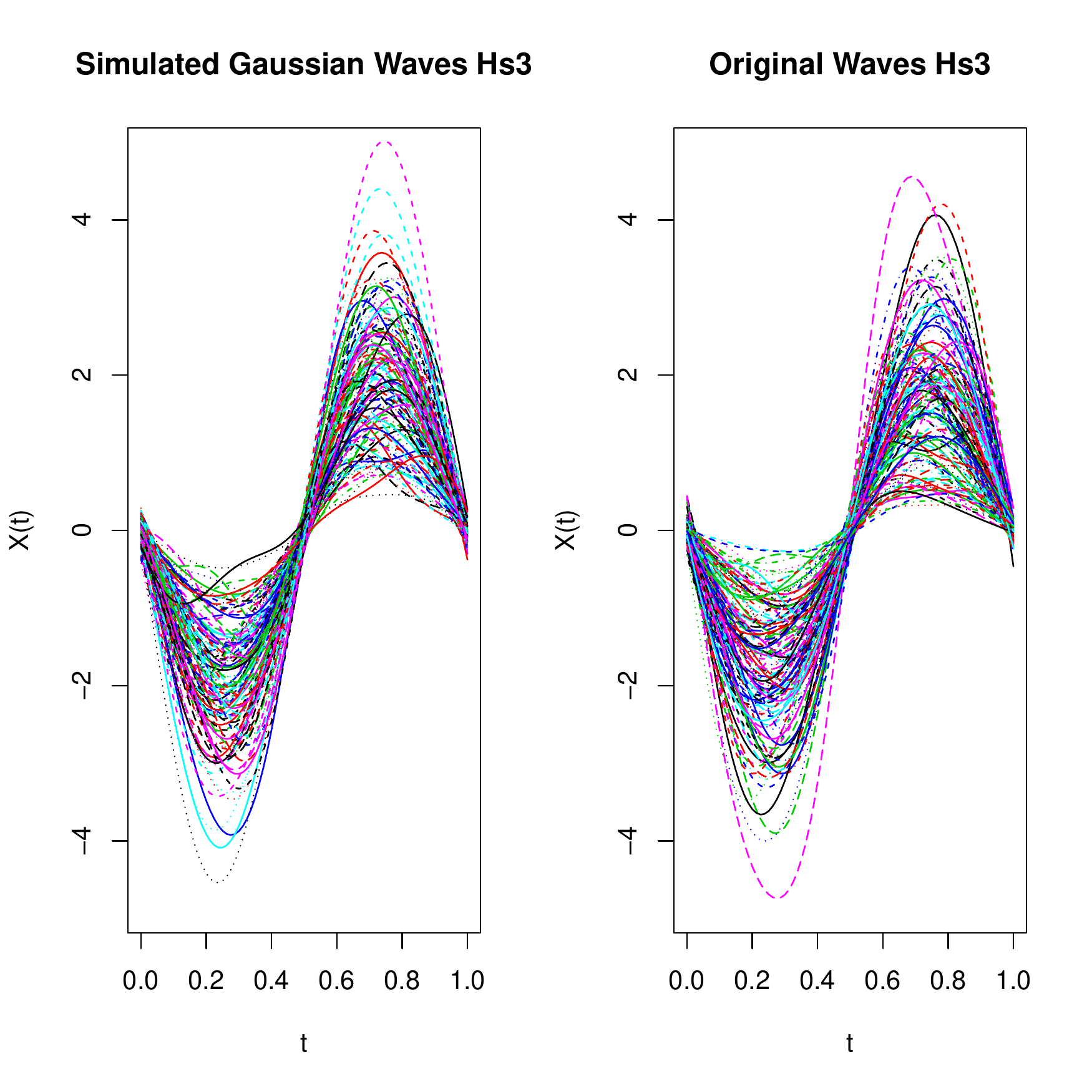} \quad
  \includegraphics[width=5cm,height=4cm]{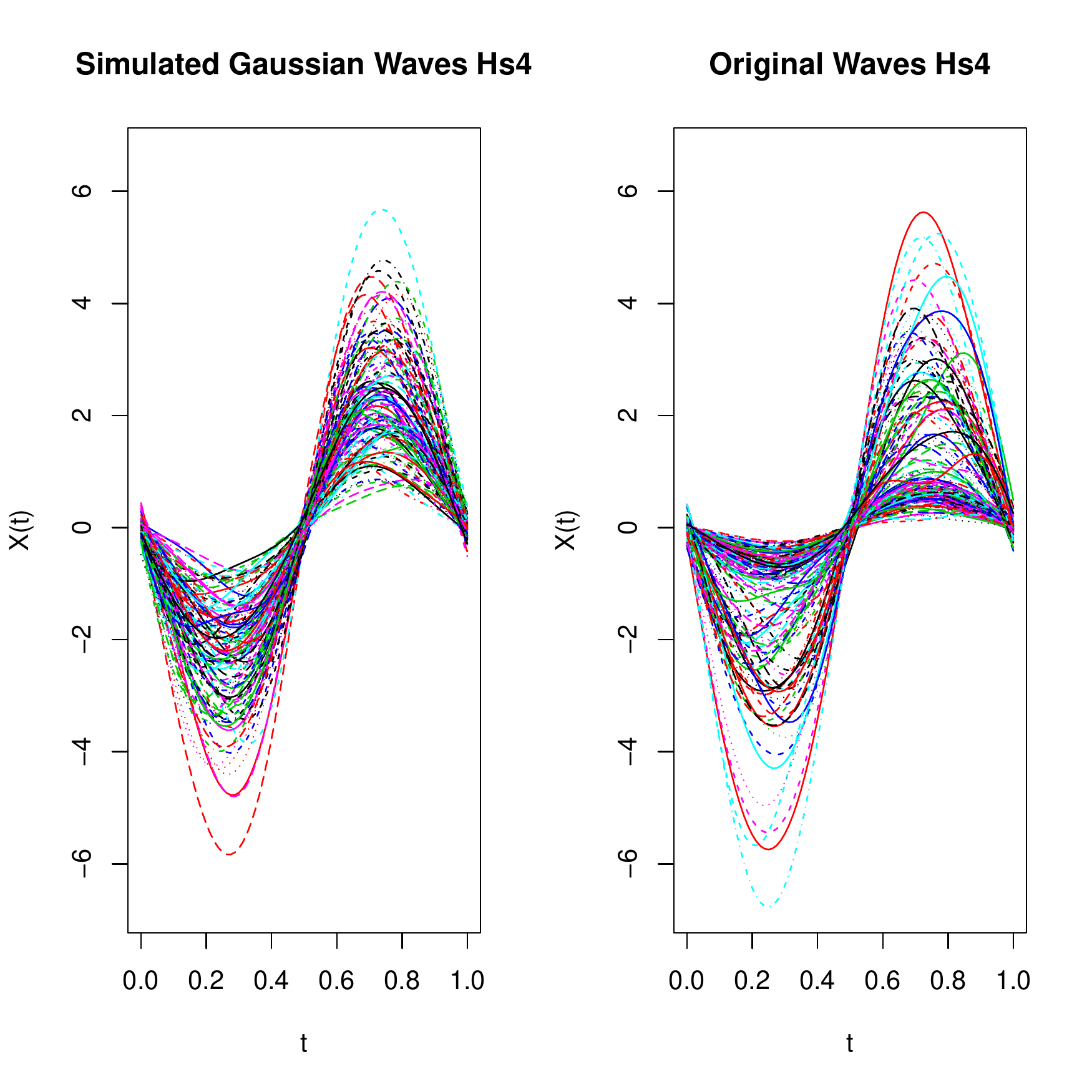}
  \caption{Real and simulated waves for the four groups considered.}\label{fig4y7}
\end{figure}

Figure \ref{fig4y7} shows the registered real waves and simulated
waves considered for this symmetry test in groups $H_s1$ to $H_s4$.
At first sight, in none of the cases is the asymmetry of the real
waves evident.

Application of the procedure described above to the selected wave
samples produced the results presented in Table \ref{tab5}. In this
analysis the $p$-values are computed with the chi-square
distribution with one degree of freedom, since a single function is
used in the vector $\tvg$. At the first two energy levels, we find
no evidence of the asymmetry, and in this sense, the mathematical
model used seems to be producing waves similar to the real ones. At
the higher energy levels, $H_s3$ and $H_s4$ the value of the
statistic is significant at the 5\% level, being much stronger the
evidence for the $H_s4$ data. This results seem in agreement with
what has been concluded in the literature by other means.

\begin{table}[ht]
\centering
\begin{tabular}{|c|c|c|c|} \hline
Energy level & Sample sizes (real $|$ simulated) & $Q_n$ value &
$p$-value \\  \hline
$H_s1$  & 85 $|$ 78  & 2.69 & 0.101 \\
$H_s2$  & 104 $|$ 156  & 0.31 & 0.578 \\
$H_s3$  & 136 $|$ 127 & 4.87 & 0.027  \\
$H_s4$  & 151 $|$ 121 & 51.58 & 6.9 $\times 10^{-13}$\\
\hline
\end{tabular}
\caption{$Q_n$ values for asymmetry test}\label{tab5}
\end{table}

As a final remark, we conclude that the proposed methodology is a
flexible tool that can be used to test the validity of different
hypothesis on functional data sets.

\section{Acknowledgements} The software WAFO \citep{wafo} developed
by the Wafo group at Lund University of Technology, Sweden,
available at http://www.maths.lth.se/matstat/wafo was used for the
calculation of all Fourier spectra and associated spectral
characteristics as well as for the simulation of Gaussian random
waves. The data for station 106 were furnished by the Coastal Data
Information Program (CDIP), Integrative Oceanographic Division,
operated by the Scripps Institution of Oceanography, under the
sponsorship of the U.S. Army Corps of Engineers and the California
Department of Boating and Waterways (http://cdip.ucsd.edu/). This
work was partially supported by CONACYT, Mexico, Proyecto An\'alisis
Estad\'{\i}stico de Olas Marinas, Fase II. It was finished while
J.O. was visiting, on sabbatical leave from CIMAT and with support
from CONACYT, M\'exico, the Departamento de Estad\'{\i}stica e I.O.,
Universidad de Valladolid. Their hospitality and support is
gratefully acknowledged.


\begin{thebibliography}{1}
%
\bibitem[Benko et al.(2009)]{benko}
Benko, M., H\"ardle, W. and Kneip, A. (2009).
\newblock Common functional principal components.
\newblock \textit{Annals of Statistics} \textbf{37}: 1-34.

\bibitem[Borgman(1972)]{borgman}
Borgman, Leon E. (1972).
\newblock {\em Statistical Models for Ocean Waves and Wave Forces},
\newblock In:  Advances in Hydroscience Vol. 8, Ven Te
Chow (Ed.), Academic Press, New York.

\bibitem[Bosq(2000)]{bosq}
Bosq, D. (2000).
\newblock {\em Linear Processes in Function Spaces.}
\newblock Lecture Notes in Statistics, Vol. 149, Springer, New York.

\bibitem[Brodtkorb et al.(2000)]{wafo}
Brodtkorb,P.A., Johannesson, P., Lindgren, G., Rychlik, I.,
Ryd\'en,  E. and E. Sj\"o (2000)
\newblock {\em {WAFO} - a {M}atlab toolbox for analysis of random waves and loads.}
\newblock In: Proc. 10th Int. Offshore and Polar Eng. Conf. (ISOPE). Vol. III, 343--350, Seattle,
USA.

\bibitem[Dudley(1987)]{dudc}
Dudley, R.M. (1987)
\newblock Universal Donsker Classes and Metric Entropy.
\newblock {\it Annals of Probability}, {\bf 15}: 1306--1326.

\bibitem[Ferraty(2011)]{ferraty}
Ferraty, F. (Editor) (2011)
\newblock {\em Recent Advances in Functional Data Analysis and Related Topics.}
\newblock Physica Verlag, Berlin.

\bibitem[Ferraty and Vieu(2006)]{fv}
Ferraty, F. and Vieu, P. (2006)
\newblock {\em Nonparametric Functional Data Analysis: Theory and Practice.}
\newblock Springer, New York.

\bibitem[Gorrostieta et al.(2014)]{goros}
Gorrostieta, C., Ortega J., Quiroz, A. J. and Smith, G. H. (2014)
\newblock Characterization of storm wave asymmetries with functional data
analysis.
\newblock {\it Environmental and Ecological Statistics} {\bf 21}(2):
263-283.

\bibitem[Hall and Van Keilegom(2007)]{hvk}
Hall, P. and Van Keilegom, I. (2007).
\newblock Two sample tests in functional data analysis starting from discrete data.
\newblock \textit{Statistica Sinica} \textbf{17}: 1511-1531.

\bibitem[Horv\'ath and Kokoszka(2009)]{hk2}
Horv\'ath, L. and Kokoszka, P. (2009).
\newblock Two Sample Inference in Functional Linear Models.
\newblock \textit{Canadian Journal of Statistics}
\textbf{37}: 571-591.

\bibitem[Horv\'ath and Kokoszka(2012)]{hk}
Horv\'ath, L. and Kokoszka, P. (2012).
\newblock {\em Inference for Functional Data with Applications}.
\newblock Springer, New York.

\bibitem[Longuet-Higgins(1956)]{lh1}
Longuet-Higgins, M. (1956).
\newblock Statistical properties of a moving wave form.
\newblock \textit{Proc. Cambridge Philosophical Society} \textbf{52}
Part 2:234-245–.

\bibitem[Longuet-Higgins(1957)]{lh2}
Longuet-Higgins, M. (1957).
\newblock The statistical analysis of a random moving surface.
\newblock \textit{ Philosophical Transactions of the Royal
Society London, Series A} \textbf{249}(966):321–-387.

\bibitem[Mu\~noz Maldonado et al.(2002)]{munoz}
Mu\~noz Maldonado, Y., Staniswalis, J.G., Irwin, L.N. \& Byers, D.
(2002).
\newblock A similarity analysis of curves.
\newblock \textit{Canadian Journal of Statistics} \textbf{30}: 373-381.

\bibitem[Ochi(1998)]{ochi}
Ochi, M. K. (1998).
\newblock {\em Ocean Waves: The Stochastic Approach}.
\newblock Cambridge Ocean Technology Series. Cambridge University Press,
Cambridge.

\bibitem[Paparoditis and Sapatinas(2014)]{ps}
Paparoditis, E. \& Sapatinas, Th. (2014).
\newblock {\em Bootstrap-based testing for functional data.}
\newblock arXiv:1409.4317v1 [math.ST].

\bibitem[Pe\~na(2012)]{pena}
Pe\~na, J. (2012).
\newblock {\em Propuestas para el problema de dos muestras con datos
funcionales.}
\newblock Tesis de maestr\'{\i}a, Universidad de Los Andes, Colombia.

\bibitem[Pierson(1955)]{pi}
Pierson, W. J. Jr. (1955).
\newblock Wind-generated gravity waves.
\newblock \textit{Advan. Geophys.} \textbf{2}: 93-178.

\bibitem[Pollard(1982)]{poll82}
Pollard, D. (1982)
\newblock A Central Limit Theorem for Empirical Processes.
\newblock {\it Journal of the Australian Mathematical Society,
Series A}, {\bf 33}: 235-248.

\bibitem[Pollard(1984)]{poll}
Pollard, D. (1984)
\newblock {\em Convergence of Stochastic Processes}.
\newblock Springer, New York.

\bibitem[Ramsay and Silverman(2002)]{rs1}
Ramsay, J.O. \& Silverman, B.W. (2002).
\newblock {\em Applied Functional Data Analysis.}
\newblock Springer, New York.

\bibitem[Ramsay and Silverman(2005)]{rs2}
Ramsay, J.O. \& Silverman, B.W. (2005).
\newblock {\em Functional Data Analysis. 2nd. Edition}
\newblock Springer, New York.

\bibitem[Torsethaugen(1993)]{Torset1}
Torsethaugen, K. (1993)
\newblock A two-peak wave spectrum model.
\newblock In {\it Proc. 18th. Int. Conference on Ocean, Offshore and Artic
  Engineering (OMAE)}, Vol~II, 175--180.

\bibitem[Torsethaugen and Haver(2004)]{Torset2}
Torsethaugen, K. and Haver, S. (2004).
\newblock Simplified double peak spectral model for ocean waves.
\newblock In {\it Proc. 14th. Int. Offshore and Polar Engineering Conference}, 23--28.

\bibitem[van der Vaart(1996)]{vdv2}
van der Vaart, Aad (1996)
\newblock New Donsker Classes.
\newblock {\it Annals of Probability}, {\bf 24}: 2128-2140.

\bibitem[van der Vaart(1998)]{vdv}
van der Vaart, Aad (1998)
\newblock {\em Asymptotic Statistics}.
\newblock Cambridge University Press, Cambridge.

\bibitem[van der Vaart and Wellner(1996)]{vdvw}
van der Vaart, A. W. and Wellner, J. A. (1996)
\newblock {\em Weak Convergence and Empirical Processes}.
\newblock Springer Series in Statistics. Springer,
New York.

%
\end{thebibliography}


\end{document}